\documentstyle[11pt,aasms4]{article}

\begin{document}

\title{The Evolution Of A Primordial Galactic Magnetic Field\altaffilmark{5}}
\author{Armando M. Howard\altaffilmark{1,2}, Russell M. Kulsrud\altaffilmark{3,4}}
\altaffiltext{1}
{Goddard Institute for Space Studies, NASA,
2880 Broadway, New York, 10025}
\altaffiltext{2}
{email:acamh@nasa.giss.gov}
\altaffiltext{3}
{Princeton University Observatory, Princeton, NJ 08544}
\altaffiltext{4}
{e-mail: rkulsrud@astro.princeton.edu}
\altaffiltext{5}
{Submitted to the Astrophysical Journal}
\begin{abstract}

	We consider the hypothesis that galactic magnetic fields  are primordial.
We also discuss the various objections to this hypothesis.  We assume that there 
was a magnetic field present in the galactic plasma before the galaxy formed.  
After the galactic  disk formed, the lines of force thread through it and  remain
 connected  to the external cosmic medium.  They enter through one side of the 
disk, proceed horizontally a distance $ l $ in the disk, and then leave through 
the other side. We find that the lines of force are stretched  rotation of the 
galactic disk, which amplifies the toroidal component of the field and increases
$l $.  When the magnetic field is strong enough, it produces ambipolar velocities
that try to lift the line out of the galactic disk but in opposite directions on
different parts of the line.  The result is, instead of the line being expelled 
from the disk, its horizontal length $l $ is shortened, both in the radial, and
in the toroidal direction.  This leads to a reduction of the rate of horizontal 
stretching, and, finally, a reduction in the magnetic field strength. After a 
sufficient time, the magnetic field at all points goes through this stretching 
and reduction, and the field strength approaches a universal function of time. 
This function is slowly decreasing, and only depends on the ambipolar properties
of the interstellar medium. At any given time the magnetic field is toroidal, and
 has the same strength  almost everywhere. On the other hand, it turns out that 
its direction varies rapidly with radius, changing sign every one hundred parsecs
field dominate over that of the other. The resulting field has a net Faraday 
rotation. If such a  field were observed  with low resolution in an external gal
axy then the field would appear toroidal in between the spiral arms. The spiral 
density wave would turn it so that the lines appear to trace out the spiral arm,
 although the apparent lines really are the sum of pieces of magnetic lines as 
they cross the disk.  They do not necessarily extend very far along the arms.
We contend that this model of the magnetic
field, which arises naturally from a primordial 
origin,  can fit the observations as well
as other models for the magnetic field, such
as those arising from the mean field dynamo theory.
Finally, because the field lines are topologically
threaded through the disk they cannot be 
expelled from the disk.  This counters
the  objection against the primordial origin, 
that such  a field could not survive very long
in the galaxy.

\end{abstract}                                                                               


\section{Introduction}

	There are two schools of thought as to
the origin of the galactic magnetic field.
The first school holds that, after formation 
of the galactic disk, there was a very weak seed
field of order $ 10^{ -17} $gauss  or so, and that  this
field was amplified to its present strength by dynamo action driven
by interstellar turbulence.  The description
of how this happens is well known(Ruzmaikin, Shukorov and
Sokoloff 1988), and
is summarized by the mean field dynamo theory.

	The second school of thought concerning
the origin of the galactic field is, that it
is primordial.  That is, it is assumed that there
was a reasonably strong magnetic field present
in the plasma before it collapsed to form the 
galaxy.  It is assumed that the field was coherent in
direction and magnitude prior to galactic formation.
The actual origin  of the magnetic field is not
specified. However, it could be formed by dynamo 
action during the actual formation of the galaxy,
(Kulsrud et al 1996).

	This paper discusses the evolution of the
structure and strength of such a primordial  magnetic field 
during the life of the galaxy.  The origin of the field  
prior to formation of the galaxy   is not considered.

	It was Fermi(1949) who first proposed
that the galactic magnetic field was of primordial 
origin.  Piddington(1957, 1964, 1967, 1981)  in a number of papers
suggested
how this might actually
have happened.  However, he supposed the magnetic
field strong enough to influence
the collapse.  One of the first
problems with the primordial origin 
is the wrap up problem.  It was pointed
out by Hoyle and Ireland(1960, 1961) and by $ \hat{O}ke $ et. al.
(1964), that the field
would wrap up into a spiral 
similar to the  spiral arms in only
two or three galactic rotations.  It
was supposed that this is the natural
shape of the magnetic field lines.  If the
winding up continued for the fifty or so 
rotations of 
the galaxy, the field lines would
reverse in direction every one hundred
parsecs.  This seemed absurd.  The various attempts
to get around this problem involved 
magnetic fields which were strong enough to 
control the flow,  and also strong radial outflows.
Several  forceful arguments against
the primordial origin   were advanced by Parker.
The arguments were that  the field would be expelled from the
galaxy, either by ambipolar diffusion(Parker 1968),  or by
rapid turbulent diffusion(Parker 1973a, 1973b)

	In this paper we reexamine the problem of a
a primordial field. 
We proceed along lines first initiated by
Piddington.  Just as in his model for the 
primordial field, 
 we start with a cosmic field with lines
of force threading the galaxy.
 We further  assume
that, after the  collapse to the disk, the lines remain 
threaded through the galactic disk.
The lines enter the lower boundary
of the disk in a vertical direction,
extend a short horizontal distance in the disk, 
and then leave through the upper boundary. 
Thus, each line initially extends a finite
horizontal distance in the disk.

	 However, in contrast to Piddington's model, we 
assume the field is too weak to affect the plasma
motions, especially the rotational motion
about the galactic center. Consequently,  it
tends to be  wrapped up by the differential rotation of the
galaxy.   In addition, we  include additional
physics of  the interstellar medium in which
the magnetic field evolves. 
After  toroidal stretching
strengthens  the
toroidal field sufficiently, 
 a  strong vertical force is exerted 
on the ionized part of the interstellar medium forcing
it through the neutral part.  We contend
that this force will not expel the entire lines
of force from the galactic disk,
but the resulting ambipolar velocity 
will only tend to shorten the length 
which each line spends in the disk as it 
threads through it.
In particular, ambipolar diffusion  will
decrease the radial component of the horizontal 
field and shorten its radial extent in the disk.
This will decrease  the toroidal stretching
of the field line, and as a result the magnetic
field strength will approach a
slowly decreasing saturated condition. 
Thus,  after several gigayears, the lines will end up 
extending  a longer distance in the azimuthal direction
than they  did initially, but a much shorter
distance in the radial direction. 
At any given time after the field saturates,
 the final strength will be independent
of its initial value.  In addition,  the field strength 
 will  depend only on the 
ambipolar properties of the interstellar medium.
[For earlier work see Howard (1995) and Kulsrud(1986, 1989, 1990).]

	As a consequence of ambipolar diffusion  plus
stretching, the  magnetic field will be almost entirely toroidal,
and it will have the same strength everywhere
at a fixed time.  However,  its sign will depend on the sign
of the initial radial component of the field at its
initial position.  Because of differential rotation 
the toroidal field will still vary rapidly in sign over a radial
distance of about one hundred parsecs.

	This model for the magnetic field evolution
would seem at first to leave us
with a wrap up problem and produce a field 
at variance with the observed field.  However,
the initial regions where the radial field is of one 
sign are expected to be
of different area than those of the other
sign provided that  the initial field is not exactly uniform.
As a consequence, at the present time in 
  any given region of the galaxy,  the toroidal field should
have a larger extent of one sign
than of the other, even, though the sign
varies rapidly.  If one now averages over larger
regions than the size of variation, one would see a mean magnetic 
field of one sign.  This averaging is actually performed 
 by the finite resolution of the observations of the magnetic field in our
galaxy or in  external galaxies. From the observations
one would not be aware of this rapid variation.

	If one ignored the spiral arms,  the field 
would be almost completely azimuthal.  However,
it is known that the density compression in the 
spiral density wave, 
twists the magnetic field to be parallel
to the spiral arm and increases its field strength
(Manchester 1974, Roberts and  Yuan 1970).
In observing radio emission from external 
galaxies, one  tends to  see the radiation 
mostly from the spiral arm, where cosmic
 rays are  intense and the magnetic field 
is stronger.  Because in these regions the
field is aligned along the arms, one would
naturally get  the impression that the field 
extends along the entire arm.  In  fact,
on the basis of our model, 
 one would actually 
be seeing short pieces of field lines pieced 
together as they cross the spiral arm and 
thread through the disk. The magnetic 
field would be  mainly
azimuthal in between the spiral arms,
and only as it crosses the arm
would it be twisted to align along the  arm.
Observationally the magnetic field of this model 
would  appear the same as that of a large scale
magnetic field.

	Similarly, one should see Faraday
rotation which is proportional to the amount
by which the effect of the toroidal field 
of one sign exceeds the other sign
in this region.  The
amount of rotation produced by  any one region would
be related to  difference in areas occupied
by the fields of different sign.  This  in turn,
is given by the amount by which the 
initial area at the initial position
where the radial field is of one sign exceeds 
the initial  area where the radial field is of opposite
sign.

	In Parker's argument(Parker 1973a, 1973b) concerning
the expulsion of the primordial field he
introduces the concept of turbulent mixing. 
Turbulent mixing correctly describes  the rate
at which the  
mean field will decrease by mixing. However,
it can  not change the number of lines
of force threading the disk, since this 
is fixed by their topology.  It only
gives the lines  a displacement. Since, near the
edge of the disk, the turbulent motions probably
decrease as the sources of turbulence do,
it is not expected  that the lines will be
mixed into the halo.  Also, because of
their topology the lines are not lost.
Only the length of their extent in the disk
can  be altered by turbulence.  Hence,
for our model, turbulent diffusion need
not destroy the primordial field as
Parker suggested.  Also,
blowing bubbles in the lines by cosmic ray 
pressure will always leave 
the remainder of the line behind, and the total number of lines
unaltered. Of course,  if the lines were entirely horizontal,
ambipolar diffusion could destroy the 
field,  since the lines may be lifted out of the disk
 bodily(Parker 1968). 

	Therefore,  Parker's
contention that the primordial 
field has a short life,  need not apply,
to our model.

\subsection{Our Model}

	The model which we consider is very simple.
We start with a large ball  of plasma whose mass is
equal to the galactic mass, but whose 
 radius is much larger than the current size of our 
galaxy (figure 1a).   We first assume that the 
magnetic field filling this sphere is uniform, 
and makes a finite angle $ \alpha $ with the rotation 
axis of the galaxy.

	Then we  allow the ball to collapse to
a sphere the size of the galaxy (figure 1b), and finally to a
disk the thickness of the galactic disk (figure 1c).  We
assume the first collapse is radial and uniform,
while the second collapse into the disk is linear
and one dimensional.

	During the time when  the galaxy contracts uniformly
into the disk 
along the $  z $  direction, where $ {\bf  \hat{z}} $ is in the 
direction of $ {\bf  \Omega} $,  we ignore any rotation.
 Then the resulting magnetic field
configuration is as in fig, (1c).  The horizontal
component of the magnetic field has been amplified by
the large compressional factor, while the vertical
component is unchanged, so that the resulting
field is nearly parallel to the galactic disk.
At this stage,  some lines enter the disk from the top and leave
from the top, e.g. line $ {\it a }$.  Some lines enter from
the bottom and leave through the top, e.g.  lines
 $ {\it b } $ and $ {\it  c} $.
Finally,  some lines enter from the bottom and leave through
the bottom, e.g. line $ {\it d} $. 
It turns out that lines  such as $ {\it a } $ and
$ {\it  d} $ are eventually expelled  from the disk 
by ambipolar diffusion,
so that we ignore them.  Now,
set the disk into differential rotation at time $ t = 0 $.
(If we include the rotation during collapse
the magnetic field will start to wind up
earlier. However,  the  result would be the same as though 
we were to  ignore the rotation during collapse, 
and then after the collapse let the disk rotate by an additional
amount equal to the amount by which
the disk  rotated during the collapse.
This is true 
provided that the initial field is too weak for ambipolar diffusion 
to  be important.)

\subsection{Results of the Model}

	We here summarize the conclusions
that we found from the detailed analysis of our model,
given in the body of this paper.

	Initially, after the disk forms, all the lines 
have a horizontal component that is larger than 
the vertical component by a factor equal
to the radius of the disk divided by its thickness
$ \approx R/D \approx 100 $.  This is the case if the initial
angle $ \alpha $ was of order 45 degrees
or at least not near 0 or 90 degrees.

	Now, because of  neglect of rotation
in the collapse, 
the  horizontal component of all the field lines is  in
 a single direction, the $ x $ direction say, so that 
\begin{equation}
{\bf  B } = B_i {\bf  \hat{x}} + B_i(D \tan \alpha /R ) {\bf  \hat{z} } =
B_i \cos \theta {\bf {\hat r }} - B_i \sin \theta {\bf {\hat \theta}} 
+ B_i (D \tan \alpha /R ) {\bf  \hat{z} }.
\label{eq:1} 
\end{equation}
After $ t = 0 $ the differential rotation of the 
disk stretches the radial magnetic field into the 
toroidal direction so that, following a given fluid
element whose initial angle is $ \theta_1 $,  one has
\begin{equation}
{\bf  B} = -B_i \cos \theta_1 {\bf {\hat r }}
+\left[B_i (r \frac{d \Omega }{d r} t) \cos \theta_1 -
 B_i \sin \theta_1 \right] 
{\bf {\hat \theta}} + B_i  (D \tan \alpha /R ) {\bf  \hat{z} }.
\label{eq:2} 
\end{equation}
After a few rotations the second component  dominates
the first and third components.
It is seen that the total magnetic 
field strength grows linearly with  time.  After the toroidal  magnetic
field becomes strong enough, 
 the magnetic force on the ionized part of  the
disk forces it  through the neutral component,  primarily in the 
$ z $ direction.

	Consider a single line of force.
Let us turn off the differential 
rotation for a moment.  In this case,
the $ z $ motion steepens the line of force,
but does not change the vertical field component.
This leads to a shortening of the line, both
in the radial direction and in the azimuthal 
direction (see figure 1d).  Now, let the differential rotation
continue.  Because the radial component of 
the magnetic field is reduced, the toroidal 
component increases more slowly.  Eventually, there
comes a time when the radial field is
small enough that  the shortening motions
in the azimuthal direction are  stronger 
than the stretching motion, and the azimuthal component of the  magnetic
field actually decreases even in the  presence
of differential  rotation.  From this time on, the magnetic
field strength decreases at a rate such 
that the vertical ambipolar velocity is just
enough to move the plasma a distance approximately
equal to  the
thickness of the galactic disk,  in the time $ t $,
i. e. 
\begin{equation} 
v_D t = D .
\label{eq:3} 
\end{equation}
Now,  $ v_D $ is essentially proportional  to the
average of the square of the magnetic field strength.
Therefore, for  a uniform partially ionized plasma  we have
\begin{equation}
v_D \approx \frac{1}{\rho_i \nu } \frac{B^2}{8 \pi D} ,
\label{eq:4}  
\end{equation} 
where $ D $ is the half thickness of the disk,
$ \rho_i $ is the ion mass density,  and
where $ \nu $ is the ion neutral collision rate.

	Thus, for asymptotically  long times one finds that
\begin{equation}
B \approx D \sqrt{ \nu \rho_i /t} .
\label{eq:} 
\end{equation} 
That is, the magnetic field strength
approaches a saturated time behavior
independent of its initial value.  The saturated time behavior
only depends on the ambipolar diffusion properties 
of the interstellar medium, and on the  time $ t $.
However, the time to reach saturation does
depend on the initial value of the magnetic field strength.

	The qualitative behavior is shown in figure 2,
where the dependence  of $ B $ on  time for different 
initial values is shown. 
 For a very weak initial radial component of  the field,
 saturation is not reached during a Hubble time.
However, for fields substantially larger than  the critical  initial 
field strength  for reaching saturation, the final saturated field 
 is independent of
the initial radial field. 

	In the interstellar medium,
 ambipolar diffusion is not well
modeled by diffusion through a uniform 
plasma.  In fact, the bulk of the mass of the
interstellar medium is in dark clouds, in which
 the degree of ionization is very low.  Also,
the outward magnetic force is concentrated
in the volume of the clouds.
As a result the ambipolar diffusion velocity
is more accurately given by the formula
\begin{equation}
v_D = \frac{B^2 (1 + \beta/\alpha )}
{8 \pi \rho_i \nu f D} ,
\label{eq:6} 
\end{equation}
where $ f $ is the filling factor for  the clouds,
$ \rho_i $ is the effective ion density in the clouds, 
$ \nu $ is the effective ion-neutral collision
rate in the clouds, and $ \beta/\alpha $ is the ratio of cosmic ray
pressure to magnetic pressure.

The  model  for the interstellar medium, which
we employ to study the magnetic field evolution, 
 is sketched  in figure 3. 
The magnetic field is anchored in the clouds
whose gravitational  mass holds  the magnetic
 field in the galactic disk.  
The magnetic  field lines bow up in between
the clouds pulled outward by magnetic pressure and by 
cosmic ray pressure.  The outward force 
is balanced by ion--neutral friction in 
the clouds themselves.  A  derivation  of
equation 6  based on this model 
 is given in section 5. 

	Taking plausible values for the properties of
the clouds, one finds that the critical initial 
value of the magnetic field,  for it to reach saturation
in a Hubble time 
is about $ 10^{-8} $ gauss.
Further, the saturated value of the field at
$ t = 10^{10 } $  years is estimated to be 
 2 microgauss.  At this field strength
the time to diffuse across the disk $ D/v_d \approx 3 \times 10^{ 9} $
years,  which is of order of  the lifetime of the disk.
   
	Finally, let us consider the structure of the saturated
field.  The time evolution of the  magnetic field
refers  to the field in the rotating frame.  
Thus, if we wish the toroidal field at time $ t $ and at $ \theta = 0 $
we need to know the initial radial component of the field  at 
$ \theta_1 = - \Omega(r) t $.  Thus,  assuming that the field 
reaches a  saturation value of $ B_S $, 
one has
\begin{equation}
B_{\theta } =  \pm  B_S ,
\label{eq:7} 
\end{equation}
where the sign is plus or minus according to the sign of
\begin{equation}
 [\cos(-\Omega(r) t )] .
\label{eq:8}  
\end{equation}
Since $ \Omega $ depends on $ r $, we see that $ B_{\theta} $ 
changes sign over a distance $ \Delta r $ such that 
$ (\Delta  \Omega)  t = ( \Delta r d \Omega/dr) t \approx \pi $, i.e. 
\begin{equation}
\frac{\Delta r}{r}  = \frac{-\pi }{r |d \Omega/dr| t} \approx 
\frac{\pi}{|\Omega | t} .
\label{eq:9} 
\end{equation}
Taking the rotation period of the galaxy  to be 
200 million years, one finds for  $ t = 10^{ 10} $ years,
$ \Delta r \approx 100 $ parsecs.

	Therefore, because the field saturates
to a constant field strength, it is predicted that a
uniform initial field, that has a  value greater
than $ > 10^{-8} $ gauss, immediately 
  after the  collapse to the disk,  leads to a 
toroidal field that varies with $ r $, at fixed $ \theta $, 
 as a square wave with a reversal 
in sign every $ 100 $ parsecs or so.  Such a field  structure 
would produce no
net Faraday rotation and would  contradict observations.

	However, this result can be traced to
our assumption that the initial magnetic field in
figure 1 was entirely uniform.  Suppose it were not uniform.
Then, after collapse, the radial field is not
 purely sinusoidal  (see figure 4).
In fact, one expects a behavior more like
\begin{equation}
B_r(t=0) = B_i[\cos \theta + \epsilon   \cos 2 \theta ] ,
\label{eq:10} 
\end{equation}
where $ \epsilon $ is an index of the nonuniformity.
As an example, for moderately small $ \epsilon  , 
 (\cos \theta + \epsilon  \cos 2 \theta ) $ is positive
for 
\begin{equation}
- \pi/2 + \epsilon   < \theta < \pi/2 - \epsilon ,
\label{eq:11}  
\end{equation}
and negative for 
\begin{equation}
\pi/2 - \epsilon  < \theta < 3 \pi/ 2 + \epsilon  .
\label{eq:12} 
\end{equation}

Therefore  the radial extent in which $ B_{\theta}(t) $ is negative
is larger than the radial extent in which it is positive
by a factor 
\begin{equation} 
\frac{\pi - 2 \epsilon  }{\pi + 2 \epsilon  } 
\label{eq:13} 
\end{equation}
The resulting field is  positive over a smaller range of $ r $ 
than the range  in which it is negative.  Consequently,
in such a magnetic field there would be a net  Faraday 
rotation of polarized radio sources.  
Note that it is easily  possible that the regions in which
the  radial field is initially weaker 
can end up as regions that dominate in flux
over those region that come from regions
in which the radial field  was stronger!  Now, if one averages this
field over regions much larger than $ 100 $ parsecs,
then the resulting mean field is smooth and 
axisymmetric.  This result  is contrary 
 to the generally held belief that 
a primordial field should lead to a field with bisymmetric
symmetry(Sofue et al 1986).   This result alone  shows that  a more
careful treatment of the evolution of the primordial field,
such as that discussed above, leads to quite  different
results from those commonly assumed. (However, as mentioned
above, including the effect of the
spiral density wave will produce a magnetic field lines
in the spiral arms  that will appear to have bisymmetrical
shape.)

	We emphasize  that, although the actual magnetic  field 
derived in our model is a tightly 
wound spiral,  the magnetic  field averaged over a 
moderate size scale appears to be smooth and  axisymmetric.

	A second important conclusion of our model
 is that because of their topology, the field lines 
cannot not be expelled from the  disk by ambipolar diffusion.
 The same line of force
  diffuses downward in the lower part of the
disk, and upward in the upper part of the disk.
The line must thus continue to be threaded through the disk.
(See lines $ b $ and $ c $ in figure 1c.)

	A third conclusion that can be drawn from
our model 
  is that any single line after saturation 
 has only a finite extent
in the disk.  For example, if the initial field is  a few microgauss,
then it turns out that the line only extends a radian
or so before leaving the disk. This finite extent of
the magnetic lines  would make  the escape 
of cosmic rays from the galaxy possible without
the necessity of 
any disconnection of the magnetic field lines in the
interstellar medium.

	A final conclusion that should be noted  is, that
the saturated  magnetic field of our model has 
 a  mean field strength  smaller than the rms field strength.
This has  the consequence  that  different 
methods for measuring the magnitude of the magnetic field
strength should lead to different results.
 The measurement by nonthermal 
radio emission of the cosmic ray electrons
measures the rms field strength while
 Faraday rotation measures the mean
strength.

	The outline of the paper is as follows:

	In section 2 an  analytic model
is developed to demonstrate the properties 
described in this introduction.

	 In section 3 a more precise one dimensional
numerical  simulation  is carried out that  confirms  the evolution
of the field in the $ z $ direction given in section 2.

	In section 4 it is shown that the three 
dimensional equations for the evolution of the field 
can be reduced to two independent variables.  These are
$ z $ and an angular coordinate $ u= \theta - \Omega(r) t $
that is constant along the spirals generated by the
differential rotation of the galaxy.
A numerical simulation of  the resulting 
differential  equations is carried out.
It is shown that after a long time the resulting 
 magnetic field does evolve locally 
in the essentially same way as is given in sections 2 and 3. 
In  addition, it varies 
in radius as a square wave with uneven lobes.

	In section 5 the astrophysics of the
interstellar medium clouds is discussed, and
a derivation is given of expression 6,
for the
effective mean ambipolar motion of the field in the disk.

	In the concluding section 6,  the implications of this
model for the evolution of a magnetic field of primordial origin  
are given.   The bearing  of these implications
 on the various
criticisms of the primordial field  hypothesis  are discussed.

\section{Local Theory: Analytic }

	In this and the next section we wish to
consider the local behavior of 
the magnetic field following a fluid element that
moves with the galactic rotation. If the ambipolar diffusion 
were  strictly in the $ z $  direction, the evolution
of the field in a given fluid element  would be
 independent of its behavior
in other fluid elements at different  $ r $ or $ \theta $.
It would   only be  affected by the differential rotation 
of the galaxy, and by ambipolar velocity in the  $ z $ direction.
 Thus, in particular, 
we could imagine the magnetic field at different
values of $ \theta $, behaving in an identical manner.  
In other words,  we could  replace the general 
 problem by an axisymmetric one.

	Let
\begin{equation} 
{\bf B } = B_r(r,z){\bf \hat{ r}} +
	   B_{\theta} (r,z){\bf \hat{ \theta }} +
	   B_z(r,z){\bf \hat{z}} .
\label{eq:14} 
\end{equation}
Let us neglect the radial velocity, and
let the ambipolar velocity be only in the $ {\bf {\hat z}} $ 
direction and proportional to the $ z $ derivative of
the magnetic field strength  squared.  Also, let us neglect any turbulent 
velocity.  The only velocities which we consider 
are the differential rotation of the galaxy,  $ \Omega(r) $,
and the ambipolar diffusion velocity of the ions.

	Then 
\begin{equation}
{\bf  v} = r \Omega(r) {\bf \hat{\theta }} + v_z {\bf \hat{z}} ,
\label{eq:15} 
\end{equation}
where 
\begin{equation}
v_z = -K \frac{\partial B^2/8 \pi }{\partial z}  ,
\label{eq:15a} 
\end{equation}
where 
\begin{equation} 
K= \frac{ (1+ \beta/\alpha)}{\rho_i f  \nu } ,
\label{eq:15b}
\end{equation}
 where as in the introduction $ \beta/\alpha $ is the ratio 
of the cosmic ray pressure to the magnetic pressure, 
$ \rho_i $ is the ion density, 
and, $ \nu $ is the ion-neutral collision rate in the 
clouds.
 The equation for the evolution of the magnetic field is
\begin{equation} 
\frac{\partial {\bf  B }}{\partial t} = \nabla \times {\bf  (v \times B }) ,
\label{eq:16} 
\end{equation} 
or, in components,
\begin{equation} 
\frac{\partial B_r}{\partial t} = \frac{\partial }{\partial z} (v_z B_r   ) ,
\label{eq:17} 
\end{equation}
\begin{equation} \frac{\partial B_{\theta}}{\partial t} =
 \frac{\partial }{\partial z}( v_z B_{\theta}) 
 + \frac{d \Omega }{dr } B_r  ,
\label{eq:18} 
\end{equation}
\begin{equation}
\frac{\partial B_z}{\partial t} = \frac{\partial v_z }{\partial r} B_r  .  
\label{eq:19}  
\end{equation}

	We  integrate these equations  numerically in section 3.
They are too difficult to handle analytically.  To treat
them approximately,  we make the assumption that for all time $ B_r $ and
$ B_{\theta} $ vary parabolically in  $ z $ i.e.
\begin{eqnarray} 
B_r =& B_r^0 (1- z^2/D^2) , \label{eq:21}  \\
 B_{\theta} =& B_{\theta}^0 (1-z^2/D^2) ,
\label{eq:22} 
\end{eqnarray}
where $ B^0_r $ and $ B^0_{\theta} $ are functions of time.
 We  apply equation~\ref{eq:17} and equation~\ref{eq:18}
at $ z = 0 $, and $ r = r_0 $, and  solve for $ B^0_r,$ and $   B^0_{\theta} $ 
as functions of $ t $.

	Thus, making use of equations 16,  18 and 19,  we find
\begin{equation}
\frac{\partial B_r}{\partial t} = - \frac{v_D}{D} B_r ,
\label{eq:23} 
\end{equation}
\begin{equation}
\frac{\partial B_{\theta} }{\partial t} =  - \frac{v_D}{D} B_{\theta} 
+ (r \frac{d \Omega }{d r})_{r_0} B_r ,
\label{eq:24} 
\end{equation}
\begin{equation}
v_D = \frac{ K}{2 \pi D} (B_r^2+ B_{\theta}^2) ,
\label{eq:25}  
\end{equation}
where everything is evaluated at $ z=0 , r= r_0$ so we drop
the superscripts on $ B_r $ and $ B_{\theta} $.
Since for galactic rotation $ r \Omega $ is essentially constant,
we have $ (r d \Omega / d r )_{r_0} = - \Omega(r_0) \equiv -\Omega_0 $.

	Initially $ B_r $ and $ B_{\theta} $ 
are of the same order of magnitude.  Also,  we expect $ B_{\theta} $ 
to grow, by stretching,  to a value much greater than its 
initial value,  before ambipolar diffusion becomes
important.  Thus, for simplicity, we make the choice of initial
conditions $ B_r = B_1 , B_{\theta}= 0 $  where  $ B_1 $ is the 
initial value for the radial component of the field.
According to equation~\ref{eq:2}, $  B_1 = B_i \cos \theta $
for a fluid element starting at $ \theta $,
for a fluid element starting at $ \theta $. 

	Now, if initially  $ v_D/D \ll \Omega $,  then 
according to equation~\ref{eq:24},  we 
expect $ B_{\theta} $ to at first grow linearly in time,
Then by equation~\ref{eq:25},  $ v_D $ increases quadratically in time
till $ v_D/D \approx \Omega $, and the ambipolar velocity 
 starts to affect the evolution
of $ B_{\theta} $  and $ B_r $.  If the initial field is small,
then the contribution of $ B_r $  to the ambipolar diffusion velocity
is never important.  In this case we can write 
\begin{equation}
v_D = v_{D1} \frac{B_{\theta}^2 }{B_1^2} ,
\label{eq:26} 
\end{equation}
where $ v_{D1}=K B^2_1/(2 \pi D) $. 
The  solution to the differential equation~\ref{eq:23}
and equation~\ref{eq:24}
with $ v_D $ given by equation~\ref{eq:26} is
\begin{equation} 
B_r =  \frac{B_1}{(1 + \frac{2 v_{D1} }{3 D} \Omega^2 t^3)^{1/2}} ,
\label{eq:27} 
\end{equation} 
\begin{equation} 
B_{\theta} = \frac{B_1 \Omega t}
{(1 + \frac{2 v_{D1} }{3 D} \Omega^2 t^3)^{1/2}} ,
\label{eq:28} 
\end{equation} 
and
\begin{equation}
v_D= \frac{v_{D1}}{D}\frac{\Omega^2 t^2}{(1 + \frac{2 v_{D1}  }
{3 D} \Omega^2 t^3)}  .
\label{eq:29} 
 \end{equation}
(The above solution, equations 28 and 29,  of equations  24, 25, and 27, 
is derived  explicitly in Appendix A.
However, it can be shown  by direct substitution  that it is  
a solution of equations 24, 25, and 27.)

	From these equations we see that for
\begin{equation}
t \ll \left( \frac{3 D}{2 v_{D1} \Omega^2} \right)^{1/3} ,
\label{eq:30}
\end{equation}
$ B_r $ is unchanged, $ B_{\theta} $ increases as $ \Omega t $,
and $ \int v_D dt \ll D $, so that up to this time ambipolar diffusion
carries the plasma only a small fraction of the disk thickness.

	On the other hand, if
\begin{equation}
t \gg \left( \frac{ 3D}{2 v_{D1} \Omega^2} \right)^{1/3} ,
\label{eq:31} 
\end{equation}
then
\begin{equation}
B_{\theta} \approx B_1 \sqrt{ \frac{3 D}{2 v_{D1} t }}
 = D \sqrt{ \frac{6 \pi }{K t} }
=  D \sqrt{ \frac{6 \pi f \rho_i \nu}{(1 + \beta /\alpha ) t} } , 
\label{eq:32} 
\end{equation}
and $ B_{\theta} $  is independent of both the initial
value of $ B_r $ and of $ \Omega $.  It depends only
on the ambipolar diffusion properties of the
interstellar medium.

	For all $ t  $,  we  have 
\begin{equation}
\frac{B_r}{B_{\theta} } = \frac{1}{\Omega t} ,
\label{eq:33} 
\end{equation}
so  for   $ t = 10^{ 10 } $  years, $ B_r = B_{\theta}/300 $,  
and the field becomes strongly toroidal.

	The question arises as to how strong 
$ B_1 $ must be in order that the saturated solution, 
equation~\ref{eq:32},  is reached.  Taking $ t = t_H $,  the 
Hubble time,  and making use of the expression for $  v_{D1} $,
one finds from equation~\ref{eq:31} that for saturation 
to be reached we must have 
\begin{equation}
B_1 > \frac{D}{t_H^{3/2}}\sqrt{ \frac{4 \pi f \rho_i \nu}
{3 ( 1+ \beta/\alpha ) }} .
\label{eq:34} 
\end{equation}

	Making use of the numbers derived in
section 5 one finds that if $ B_{\theta} $ is saturated 
at $ t = t_H $, then
\begin{equation}
B_{\theta} = \frac{D}{100 \mbox{pc} }
\left( \frac{10^{ 10} \mbox{years} }{t_h} \right)^{1/2}
\left( \frac{6 \times 10^{ -4} }{n_i n_0} \right) ^{1/2}
1.9 \times 10^{ -6} \mbox{gauss} ,
\label{eq:36}
\end{equation} 
where $ n_i $ is the ion density in the clouds, assumed to 
be ionized carbon, and $ n_0 $ is the mean hydrogen
density in the interstellar medium. The densities
are in cgs units. 

	 For saturation,  the critical value for $ B_1 $ is
\begin{equation}
B_{1crit} =  \frac{D}{100 \mbox{pc} }
\left( \frac{10^{ 10} \mbox{years} }{t_h} \right)^{3/2}
\left( \frac{6 \times 10^{ -4} }{n_i n_0} \right) ^{1/2}
4 \times  10^{ -9} \mbox{gauss} 
\label{eq:37} 
\end{equation}
where the densities are in cgs units.

	Hence, for the above properties of the clouds
and for an initial radial field  greater than the critical
value, the magnetic field at $ t = t_H $ saturates
at about the presently observed value.  Such
 a field arises from compression if, when the galaxy was a sphere
of radius $ 10 $ kiloparsecs, the magnetic field strength
was greater than $ 10^{ -10} $ gauss.  If before this
the virialized radius of the protosphere was $ 100 $ 
kiloparsecs, then in this sphere  the comoving initial value for the
cosmic field strength had to be greater 
than $ 10^{ -12 } $ gauss.  (That is if the cosmic field filled
all space then it had to be so strong that the present
value of the magnetic field in intergalactic space
must now be $ 10^{ -12} $ gauss.)   There are good reasons to
believe that  a magnetic field stronger than this minimum 
value 
could have been generated 
by turbulence in the protogalaxy during
its collapse to this virialized radius (Kulsrud et al. 1996). 
However, this field would
be local to the galaxy and not fill all space. 

	Finally,  one expects that  $ B_z \approx B_1/100 $,
from the initial compression into the disk.  Taking $ B_z $ 
as unaffected by the differential stretching
and by the ambipolar diffusion velocity, one
can derive an expression for the length of a line of force:
\begin{equation} 
\frac{r d \theta }{d z } = \frac{B_{\theta}   }{B_z} =
\frac{100 \Omega t}
{(  1 + \frac{2 v_{D1} }{3 D} \Omega^2 t^3 )^{1/2} }
\approx 100 \sqrt{ \frac{3 D}{2 v_{D1} t}} ,
\label{eq:38} 
\end{equation}
or
 \begin{equation}
r \Delta \theta = 
\left( \frac{ (2.5  \times 10^{-6} \mbox{gauss}   }{B_1} \right) 10
 \mbox{kpc} .
\label{eq:39} 
\end{equation}

	Thus, for example  if $  B_1 = 4.5 \times 10^{ -7} $ gauss, then
a line of force passing through the solar position
in the galaxy,  would stretch 
once around the galaxy.  Stronger initial fields lead to
shorter lines of force.  If the initial field is stronger
than 2 microgauss,  then it is possible that cosmic rays
 can escape along the lines of force into the halo during
their average life time.

\bigskip

\section{One Dimensional Numerical Simulation}

	It is clear that any gradient of the magnetic
field strength will lead to ambipolar diffusion.
However, gradients in the angular direction
are weak, so that  we may neglect ambipolar diffusion in the
angular direction.  Similarly, the gradients in the radial
direction are weak at first, although as pointed out in
the introduction, the magnetic field ends up reversing rapidly
in the toroidal direction, so eventually 
radial ambipolar diffusion becomes as important as vertical
ambipolar diffusion.

	In this section, we restrict ourselves
to gradients only  in the $ {\bf \hat{z}} $  direction.
In this case, the axisymmetric approximation is
valid, and  the relevant equations  are equation~\ref{eq:17}
and equation~\ref{eq:18},  where 
$ v_z $ is given by equation~\ref{eq:15a},
\begin{equation}
v_z = - K \frac{\partial }{\partial z} 
\frac{B^2}{8 \pi}  .
\label{eq:40}  
\end{equation} 

	In the previous section, these equations were
reduced to zero dimensions by the {\it ansatz} 
that $ B_r $ and  $ B_z $ were parabolic in $ z $ according to 
equation~\ref{eq:21} and equation~\ref{eq:22},
and the basic   equations were applied only at
$ z = 0 $.  In the present  section we treat these equations 
numerically for all $ z $ with $ |z| < D $,
and drop the parabolic assumption.  
We treat the disk as uniform so that $ \rho_i $ and $ \nu $
are taken as constants.

	We need boundary conditions at $ z = \pm D $.
We expect $ B $ to be quite small outside the disk,
$ |z| > D $. (This is because we suppose that 
 neutrals are  absent in the halo
and $ \nu = 0 $,  so that the flow velocity $ v_z $ becomes very
large.  Since  flux is conserved, $ v B $ must be a constant
in a steady state and 
 the magnetic field must
be very small.)  In order to match 
smoothly to the outer region, we first assume that $ \rho_i $
is a constant for all $ z $ and is very small.
In addition, we assume that $ \nu $ is constant for
$ |z| < D $, and decreases rapidly to zero in  
a narrow region,  $ D <  |z| < D + \Delta $.  Then because
$ \nu \approx  0 $ in the halo region, $ |z| > D $, 
$ v_z $ becomes so  large  that inertia is important.

	The equation for $ v_z $ should read
\begin{equation}
\rho_i \left(  \frac{\partial v_z }{\partial t}
+ v_z \frac{\partial v_z }{\partial z}\right)   = 
- \frac{\partial B^2/ 8 \pi }{\partial z}
- \rho_i \nu'  v_z ,
\label{eq:41}
\end{equation}
where we have set $ \rho_i \nu' = 
(\rho_i \nu/f) (1 + \beta / \alpha )/f $. 
Because  the time evolution is over the Hubble time $ t_H $, 
 $ \partial v_z / \partial t \ll v_z \partial v_z / \partial z $,
i.e. 
$ v_z /D$  is very  large compared to $ 1/t_H $.
Thus, we drop the partial time derivative.

	The equations for $ B_r $ and $ B_{\theta} $ are the same as 
in the disk.  However, because $ v_z $ is large 
 the divergence  terms $ \frac{\partial }{\partial z }(v_z B_r) $
and $ \frac{\partial }{\partial z }(v_z B_\theta ) $ are much 
larger than the other terms in the region
near $ |z| = D $,  and in the halo.  Thus, we have
\begin{equation}
 \frac{\partial }{\partial z }(v_z B)  = 0  ,
\label{eq:42} 
\end{equation}
where $ B = \sqrt{ B^2_r + B_{\theta}^2 } $  is the magnitude
of the magnetic field.  This means that 
\begin{equation}
v B = \Phi = \mbox{const} ,
\label{eq:43} 
\end{equation}
where $ \Phi $ is the rate of flow of magnetic flux,
and it is a constant in these regions.  (We drop the 
subscript on $ v $)

	Dividing equation 41  by $ \rho_i $ and dropping
the partial time  derivative, we get
\begin{equation}
\frac{\partial }{\partial z} \left( v^2 + \frac{B^2}{4 \pi \rho_i} \right) 
= - 2 \nu' v  .
\label{eq:44} 
\end{equation}
 Combining this equation with equation~\ref{eq:43} we have
\begin{equation}
\frac{\partial }{\partial z} 
\left( v^2 + \frac{\Phi^2}{ 4 \pi \rho_i \nu'^2}  \right) 
= - 2 \nu' v  ,
\label{eq:45} 
\end{equation}
or 
\begin{equation}
\frac{d \left( v^2 + \Phi^2 / 4 \pi \rho_i v^2 \right) }{v} ,
= - 2 \nu' d z  ,
\label{eq:46} 
\end{equation}
or 
\begin{equation}
- \int \left( ( \frac{\Phi^2}{4 \pi \rho_i v^4 } - 1 \right) d v
= - \int \nu' d z  .
\label{eq:47} 
\end{equation}

	For $ z $ larger than $ D + \Delta  $,  the 
right hand side   becomes a constant,  so that 
$ v^4 \rightarrow \Phi^2/4 \pi \rho_i $.
As $ z $ decreases below $ D $,  the right hand side becomes linear in
$ z $, and we find that for $ z $  far enough into the disk,
the inertial term becomes negligible.
Hence, 
\begin{equation} 
3 v^3 \approx \frac{\Phi^2}{4 \pi \rho_i \nu'  (D - z)} ,
\label{eq:48} 
\end{equation}
and $ v $ approaches a small value.  This result 
breaks down when $ v $ is small enough that
the other terms in equation~\ref{eq:17} and equation~\ref{eq:18} 
become important.

	Thus, the connection of the main part of the disk
to the halo is given by equation~\ref{eq:47}
\begin{equation}
\int^{v_c}_v \left( \frac{\Phi^2}{4 \pi \rho_i v^4} - 1 \right) d v
= - \int^z_{z_c} \nu' d z  ,
\label{eq:49} 
\end{equation}
where $ v_c = (\Phi^2/ 4 \pi \rho_i)^{1/4} $.  Transforming 
this  to an equation for $ B $ we have
\begin{equation}
\int^B_{B_c} \left( \frac{B^4}{4 \pi \rho_i \Phi^2 } - 1 \right) 
\frac{\Phi d B}{B^2} = - \int^z_{z_c} \nu' d z ,
\label{eq:50} 
\end{equation}
where $ B_c = \Phi/v_c = (4 \pi \rho_i \Phi^2 )^{-1/4} $,  and where
$ z_c $ is the value of $  z $ where $ B = B_c $.
Let us write $ \int^{\infty }_D \nu' dz = \nu'_0 \Delta' $,  where
on the right hand side $ \nu'_0 $ denotes the constant 
value of $ \nu'  $ in the disk.  Then for
$ z $ far enough into the disk,  equation~\ref{eq:50}  reduces
to
\begin{equation}
\frac{B^3}{3 ( 4 \pi \rho_i)} \approx  (D - \epsilon - z) \nu_0 ,
\label{eq:51}
\end{equation}
where
\begin{equation} 
\epsilon = \Delta' + 2 v_c/ 3 \nu_0  .
\label{eq:52} 
\end{equation}

	Now, $ \Delta' $ is small by assumption.  If we estimate
$ \Phi $ as $ B_0 v_{D \mbox{eff}} $,    where $  v_{D \mbox{eff}} $ 
is the order of the ambipolar diffusion velocity at the 
center of the the disk, then $ v_c \approx (B_0 v_{D \mbox{eff}} /
\sqrt{ 4 \pi \rho_i } )^{1/2} \ll B_0/\sqrt{ 4 \pi \rho_i } $,
so that $ \epsilon $  is much smaller than the distance an 
alfven wave can propagate in the ion-neutral collision time.
This is clearly a microscopic distance compared to the
thickness of the disk,  so that we may neglect it.

	In summary, equation~\ref{eq:50} 
implies that the inner solution
essentially vanishes at $ z \approx \pm D $, so that we may take our boundary
condition to be $ B_r = B_{\theta} = 0 $ at $ z = D $.
Given this solution the above analysis shows how to smoothly continue 
it into the halo.

	 Equation~\ref{eq:17} and equation~\ref{eq:18}
can be made dimensionless by a proper transformation.
 We choose a transformation that is consistent
with that employed in the next section. 
 Define a unit of time $ t_0 $ by
\begin{equation}
\Omega_0 t_0 = R/D  ,
\end{equation}
where $ R $ is the radius of the sun's galactic orbit,
and $ \Omega_0 = \Omega(R) $ is its angular velocity.
Next, choose a unit of magnetic field $ B_0 $ to satisfy
\begin{equation}
\frac{K B_0^2}{4 \pi D} = \frac{D}{t_0} .
 \label{eq:54}  
\end{equation}
For such a field the ambipolar diffusion velocity is
such that the field lines cross  the disk in a time $ t_0 $.
Now, let 
\begin{eqnarray} 
t &= & t_0 t'  , \\  \nonumber 
B_r &= & (D/R) B_0 B'_r   ,\\ \nonumber 
B_{\theta } &= & B_0 B'_{\theta}  ,\\  \nonumber 
z &=& D z'  .
\end{eqnarray}

	For a cloud  ion density of $6 \times  10^{-4}/ \mbox{cm}^3 $,
 and a mean interstellar medium density of $  n_0 = 10/ \mbox{cm}^3  ,
B_0 = 2.85  \times 10^{ -6} \mbox{gauss} $, (see section 5).
  Also, $ D = 100 $ parsecs, and 
$ t_0 = 3 \times 10^{ 9} $ years.  The Hubble time in these dimensionless 
units is about 3.  

	The details of the numerical simulation are
given in another paper, (Howard, 1996).  The dimensionless
equations to be solved are
\begin{equation}
\frac{\partial B'_r}{\partial t'} =  \frac{1}{2}  \frac{\partial }{\partial z' }
(\frac{\partial B'^2}{\partial z'} B'_r)  ,
\label{eq:55} 
\end{equation}
\begin{equation}
\frac{\partial B'_{\theta} }{\partial t'} =  \frac{1}{2 } \frac{\partial }{\partial z' } 
(\frac{\partial B'^2}{\partial z'} B'_{\theta} ) -  B'_r   .
\label{eq:56} 
\end{equation} 
 Our dimensionless boundary condition is $ B'_r = B'_{\theta} = 0 $
at $ z = \pm 1 $.

	The results of the integration of these equations are shown 
in figures 5 and 6. The initial profiles  of $ B_r $ and $ B_{\theta} $ 
are parabolic. 

	In figure 5 
the initial value of $ B'_r $ is $ B_i = 0.01 $,  so that $ B_r =
2.85 \times 10^{ -8 }  $  gauss.  It is seen that the qualitative
behavior is the same as that described in sections 1 and 2.  The
field at first grows and then relaxes back to a decaying solution.
Figure 6 gives the time evolution of $ B'_{\theta} $
at $ z= 0 $ for the set of   initial conditions, $ B_r = B_i \cos theta $ 
where $ \theta $ runs from zero to one hundred eight degrees
in increments of fifteen degrees. 
It is seen that the curves all tend to the
same asymptotic behavior and are similar to those
in figure 2.  As in figure 2,   the field from weaker initial values
of $ B_r $ takes  longer times to reach its
peak value and the saturation curve.  These results
are similar to those which are
represented by equations 28 and 29.

 It is seen in figure 5 that, except near the edge, the profile
remains similar to a parabola.  Near the edge the cube root
behavior of equation~\ref{eq:51} is also evident.  This cube root
behavior can be understood since the flux 
$ v_z B $ is roughly constant, and therefore $ 
( \partial B'^2 / \partial z') B' \approx
 (\partial/\partial z' ) (1 - z')^{2/3} \times 
(1-z')^{1/3}  \approx (1-z')^0 =$ a constant.

	 The most important
feature in  figure 5 and 6, is the saturation of 
$ B $ to the envelope curve.
This saturation occurs if the initial value of $ B'_r  $ is
larger than   0.01.  Thus, even
in our more precise calculations, information on
the initial B is lost, and the final value of $ B $
at fixed time depends only on the initial sign of $ B_r $.

\section{Two Dimensional Analysis}

	In the last section we have treated the
evolution of the magnetic field under the
influence of differential rotation and ambipolar diffusion 
in a one dimensional approximation.  Only
the $ z $ component of the ambipolar diffusion was kept.
In this approximation 
the magnetic field in each column of fluid at $ r, \theta $
evolves independently of any other $ r , \theta $ 
column of fluid.  

	In early times, because the
galactic disk is thin compared to its radius,
this is a good approximation, since  the
horizontal gradients of $ B^2 $ are small
compared to the vertical gradients.
However, because of the differential 
rotation of the galaxy, fluid elements
at different radii, that were initially very far from each other
are brought much closer together, and the
horizontal gradients are increased until
they become comparable to the vertical gradients.

	For example, two fluid elements that
were initially on opposite sides of the galaxy, and
at a difference in radius of one hundred
parsecs, will be brought to a position on
the same radius after about fifty rotations.
Since the evolution of the field in these
two fluid elements is very different, we expect
the gradient of $ B^2 $ in the radial
direction to become as large as that in the
vertical  direction.   As a consequence, ambipolar drift
velocities in the horizontal direction can be
expected to be as large as those in the
vertical direction.  On the other hand,  two
fluid elements at the same radius, but initially 
far apart, remain far apart.   The 
ambipolar motions in the $ \theta $ direction
should remain small.

	Thus, we expect that, at first, only ambipolar 
$ z $ motions are important for the evolution of the magnetic field,
but that eventually the radial ambipolar  motions also 
will become  important, although not the angular ambipolar  motions.
That is, we expect the problem to become 
two dimensional.

	In order to properly demonstrate 
this evolution, we introduce a new independent variable
\begin{equation}
u \equiv \theta - \Omega(r) t .
\label{ueq}
\end{equation}
If ambipolar diffusion is neglected, this variable is just the
initial angular position of a fluid element
that is at the position $ r, \theta, z $ at time $ t $.
Since in the absence of ambipolar diffusion 
the variables $ r, u, $ and $ z $ are constant,
following a given fluid element.  They
 would be  the  Lagrangian variables 
if only rotational motion is considered.
It is appropriate to describe the evolution 
of the magnetic field components $ B_r,  B_{\theta} $,
and $ B_z $ in terms of these variables.

	Inspection of equation~\ref{ueq} shows that
when $ \Omega t $ is large, $ u $ varies rapidly
with $ r $ at fixed $ \theta $, in agreement with 
the above qualitative discussion.  Changing $ r $ by only
a small amount will change the initial angular
position by $ \pi $.  Thus, we expect that the
components of $ {\bf B} $ will vary finitely with 
$ u $,  but only slowly with $ r $ for fixed $ u $.  The
surfaces of constant $ u $ are tightly wrapped spirals.
Thus, the behavior of the field should be finite in
$ r, u, $ and $ z $, but only gradients with respect to
$ u $ and $ z $ should be important.

	To see this, let us first write the total velocity,
$ {\bf  v = w } + \Omega r {\bf  \hat{\theta }} $, 
where $ {\bf  w} $ is the ambipolar velocity. Next, let us
derive the equations for the components 
$ B_r, B_{\theta} ,  B_z $ and $ w_r, w_{\theta}, w_z $ 
in terms of the Eulerian variables $ r, \theta $ and $ z $.
\begin{equation} 
\frac{\partial B_r}{\partial t} =
({\bf  B \cdot \nabla }) w_r - ({\bf  w \cdot \nabla }) B_r  
- B_r ( \nabla \cdot {\bf  w}) - 
\Omega \frac{\partial B_r }{\partial \theta }, 
\label{eq:a2} 
\end{equation}
\begin{equation}
\frac{\partial B_{\theta}}{\partial t} =
({\bf  B \cdot \nabla }) w_{\theta} - \frac{B_{\theta} w_r }{r} 
({\bf  w \cdot \nabla }) B_{\theta}  + \frac{B_r  w_{\theta} }{r} 
 - \Omega \frac{\partial B_r }{\partial \theta } 
- B_{\theta} ( \nabla \cdot {\bf  w}),
\label{eq:a3} 
\end{equation}
\begin{equation}
\frac{\partial B_z}{\partial t} =
({\bf  B \cdot \nabla }) w_z - ({\bf  w \cdot \nabla }) B_z  
 - \Omega \frac{\partial B_z }{\partial \theta } 
- B_z ( \nabla \cdot {\bf  w}), 
\label{eq:a4} 
\end{equation}
where the ambipolar velocities are:
\begin{equation}
w_r= -K \frac{\partial B^2 /8 \pi  }{\partial r} ,
 w_{\theta} =  -K \frac{\partial B^2/ 8 \pi  }{r \partial \theta } ,
w_z =  -K \frac{\partial B^2/ 8 \pi  }{\partial z},
\label{eq:a5} 
\end{equation} 
and
\begin{equation}
\nabla \cdot {\bf  w} = \frac{\partial w_r }{\partial r} 
+  \frac{\partial w_{\theta}   }{ r \partial \theta } 
+ \frac{w_r}{r}  + \frac{\partial w_z  }{\partial z} .
\label{eq:a6} 
\end{equation}
Finally, let us transform these equations to the new
coordinates $ r, u, z $.  In doing this we assume
 that the galactic rotation velocity 
$ v_c = \Omega r $ is a constant.

	The result of the transformation is
\begin{eqnarray} 
\frac{\partial B_r  }{\partial t } &= &
\frac{B_{\theta} }{r} \frac{\partial w_r }{\partial u } 
+ B_z \frac{\partial w_r }{\partial z} \\ \nonumber 
&  & -w_r \frac{\partial B_r } {\partial r} 
-\frac{w_{\theta}  }{r} \frac{\partial B_r  }{\partial u } 
- \Omega t \frac{w_r }{r} \frac{\partial B_r  }{\partial u}
 - w_z \frac{\partial B_r }{\partial z} 
 \\ \nonumber 
& &- B_r \left(  \frac{w_r }{r}
+ \frac{\partial w_{\theta}  }{r \partial u } 
+ \frac{\partial w_z  }{\partial z } \right) ,
\label{eq:a7} 
\end{eqnarray} 
\begin{eqnarray} 
\frac{\partial B_{\theta} }{\partial t}& =&
B_r  \frac{\partial w_{\theta}   }{\partial r} 
+ \frac{\Omega t}{r} B_r 
 \frac{\partial w_{\theta}  }{\partial u} \\ \nonumber 
& &+ B_z \frac{\partial w_{\theta}   }{\partial z} 
-w_r \frac{\partial B_{\theta} }{\partial r }
-  \frac{w_{\theta} }{r} \frac{\partial B_{\theta} }{\partial u} 
-\frac{\Omega t}{r} w_r \frac{\partial B_{\theta} }{\partial u}  \\ \nonumber
& &- w_z \frac{\partial B_{\theta}  }{\partial z } 
-\Omega B_r \\ \nonumber
& &- B_{\theta} \frac{\partial w_r  }{\partial r} 
-\frac{\Omega t }{r} B_{\theta} \frac{\partial w_r  }{\partial u }  
- B_{\theta} \frac{\partial w_z  }{\partial z} , 
\label{eq:a8} 
\end{eqnarray} 
\begin{eqnarray}
\frac{\partial B_z }{\partial t } &=&
B_r  \frac{\partial w_z} {\partial r }  
+ \frac{B_{\theta }  }{r} \frac{\partial w_z }{\partial u}
+ \Omega t \frac{B_r  }{r}    \frac{\partial w_z }{\partial u}\\ \nonumber 
& & - w_r \frac{\partial B_z }{\partial r }
-\frac{w_{\theta}  }{r} \frac{\partial B_z  }{\partial u } 
- \Omega t \frac{w_r }{r} \frac{\partial B_z  }{\partial  u} \\ \nonumber
& & - w_z \frac{\partial B_z }{\partial z}
- B_z \frac{\partial w_r }{\partial r}\\ \nonumber
& & -\Omega t \frac{B_z }{r} \frac{\partial w_r }{\partial u}
- B_z \frac{w_r }{r} 
- \frac{B_z }{r} \frac{\partial w_{\theta}  }{\partial u} ,
\label{eq:a9} 
\end{eqnarray}
\begin{eqnarray} 
w_r = -K \frac{\partial B^2/ 8 \pi }{\partial r }   
-K \frac{\Omega t}{r} \frac{\partial B^2/ 8 \pi }{\partial u},
\label{eq:a10} 
\end{eqnarray} 
\begin{eqnarray} 
w_{\theta}  = - \frac{K}{r} \frac{\partial B^2/ 8 \pi }{\partial u }   ,  
\label{eq:a11} 
\end{eqnarray} 
\begin{eqnarray} 
w_{z}  = - K \frac{\partial B^2/ 8 \pi }{\partial z }.   
\label{eq:a12}   
\end{eqnarray}

	Only a few of these terms are important.
To see this let us introduce dimensionless variables for
the velocity and field components. (We will denote the
dimensionless variables by primes.) 
We will choose these variables based on our 
one-dimensional results, as follows: 

	The unit of length for the $ z $ variable is the galactic 
disk thickness $ D $. 
\begin{equation}
z= D z' .
\label{eq:a13} 
\end{equation}
The variation  of quantities with $  r $  is finite over
the distance  $ R $ the radius of the sun's orbit
in the galaxy, so we set
\begin{equation} 
r = R r' .
\label{eq:a14} 
\end{equation}
The variable $ u $ is already dimensionless and quantities
vary finitely with it, so  we 
leave it unchanged.

	The unit of time $ t_0 $ should be of order
of the age of the disk.  During this time the number of
radians through which the galaxy rotates, $ \Omega t $
is of the same order of magnitude as the ratio $ R/D $, 
so for analytic convenience we choose $ t_0 $ so that
\begin{equation} 
\Omega_0 t_0 = R/D ,
\label{eq:a15} 
\end{equation} 
and set
\begin{equation}
t = t_0 t'.
\label{eq:a16} 
\end{equation}
If we take $ R/D = 100 $,  and $ \Omega_0 = 2 \pi /( 2 \times 10^{ 8}
\mbox{years}),   $ then  $ t_0 = 3 \times 10^{ 9 } $ years.

	It is natural to choose the unit for $ B_{\theta}, B_0 $ 
as that field whose $ z $  gradient produces an
ambipolar $ z $ velocity of order $ D/t_0 $, that is an average velocity
near that which would be produced by the  saturated field.
Thus, we choose $ B_0 $ so that
\begin{equation} 
K B_0^2/(4 \pi D) = D/t_0 ,
\label{eq:a17} 
\end{equation}
and set 
\begin{equation}
B_{\theta} = B_0 B'_{\theta} .
\label{eq:a18} 
\end{equation}
(Note that the definitions in these units
are consistent with those of section 3.)

	In most cases of interest to us,
 $ B_r  $ and  $ B_z $ are much smaller
than $ B_0 $. ( In general   $ B_{\theta} $ is  initially 
small compared to $ B_0 $, but it  grows by a factor of $ R/D $ to 
up to the saturated value of  about $ B_0 $.)  Thus, we set 
\begin{equation}
B_r  = (D/R) B_0 B'_r  .
\label{eq:a19} 
\end{equation}
The vertical field $ B_z $ starts out even weaker
than this magnitude 
since the initial horizontal components of the
magnetic field were amplified by the initial compression
which formed the disk.  However, the
$ B_z $ field is amplified up to the size of the
$ B_r  $ field by the shear of the  radial ambipolar velocity
acting on $ B_r $ .
We thus transform $ B_z $ by
\begin{equation}
B_z = (D/R) B_0 B'_z  .
\label{eq:a20} 
\end{equation} 

	The ambipolar velocities $ w_r,  w_{\theta}  $, and $ w_z $ 
arise from gradients of the magnetic field in the corresponding
directions.
Thus,  after $ B_{\theta} $ has been amplified by stretching
to be of order $ B_0, w_z \approx K B_0^2/D $.  Similarly,
after the differential rotation has acted to
reduce the scale of the variation of $ B^2 $ in the radial  
direction,  $ w_r $ becomes of order $ w_z $.
However, the scale of variation in the 
$ \theta $ direction remains of order $ R $ over
the age
of the galaxy so that $ w_{\theta}  \approx K B_0^2/R  $.
Thus,  we change the $ w $ components  to
$ w' $ components by 
\begin{eqnarray}
w_r  & = &(D/t_0) w'_r ,\\ \nonumber
w_{\theta}& = &(D^2/R t_0) w'_{\theta} , \\ \nonumber 
w_z & = & (D/t_0) w'_z .
\label{eq:a21} 
\end{eqnarray} 

	Now if we transform
 equations 64 to 69   by the change of variables 
 equations 70 to 78,  and clear the dimensional factors $ t_0, B_0 $ etc. 
from the left hand side, we find that the terms on the
right hand side are either independent of dimensional
units entirely,  or are proportional to powers of $ (D/R)  << 1 $.
The full equations are given in appendix B.
Dropping these ``smaller''  terms proportional to 
a power of $ D/R $ greater than zero,  we find that the 
equations for the dimensionless variables,  to lowest order in
D/R are 
\begin{eqnarray} 
\frac{\partial B'_r} {\partial t' } &=&
- \frac{1}{r'^2} w'_r t' \frac{\partial B'_r    }{\partial u }
- \frac{\partial ( w'_z B'_r  )  }{\partial z'}  \\ \nonumber
& & + \frac{ B'_{\theta} }{r'} \frac{\partial w'_r }{\partial u} 
+ B'_z \frac{\partial w'_r }{\partial z' } ,
\label{eq:a22} 
\end{eqnarray}
\begin{eqnarray}
\frac{\partial  B'_{\theta} }{\partial t'} &=&
- \frac{t'}{r'^2} \frac{\partial (w'_r  B'_{\theta} ) }{\partial u}
- \frac{\partial w'_z  B'_{\theta}  }{\partial z' } \\ \nonumber
& &- \frac{B'_r    }{r'} ,
\label{eq:a23} 
\end{eqnarray}
\begin{eqnarray}
\frac{\partial B'_z }{\partial t' }  &=&
\frac{ B'_{\theta} }{r'} \frac{\partial w'_z }{\partial u}
+ \frac{t' B'_r    }{r'^2} \frac{\partial w'_z }{\partial u}\\ \nonumber 
& &- \frac{t' w'_r }{r'^2} \frac{\partial B'_z }{\partial u} 
- w'_z \frac{\partial B'_z }{\partial z} 
- \frac{t'}{r'^2} B'_z \frac{\partial w'_r }{\partial u} ,
\label{eq:a24} 
\end{eqnarray}
\begin{equation}
w'_r = - \frac{t'}{2 r'^2} \frac{\partial  B'^2_{\theta}}{\partial u} ,
\label{eq:a25} 
\end{equation}
\begin{equation}
w'_{\theta} = - \frac{1}{2} \frac{\partial  B'^2_{\theta}}{\partial u} ,
\label{eq:a26} 
\end{equation}
\begin{equation}
w'_z = - \frac{1}{2}  \frac{\partial  B'^2_{\theta} }{\partial z' } .
\label{eq:a27} 
\end{equation}

	Note that $ w'_{\theta} $ does not occur in the equations
for the evolution     of the $ B' $ components.  Also,
the $ r' $ derivatives are absent from these lower order
equations.  The initial conditions on $ B'(t', u, z'; r' ) $ 
are 
\begin{eqnarray}
B'_r( 0, u, z', r')&=& B_r(0,r'R ,u, z'D)  , \\ \nonumber 
 B'_{\theta}( 0, u, z', r') &=&  B_{\theta}(0,r'R ,u, z'D) , \\ \nonumber 
B'_z ( 0, u, z', r') &=& B_z(0,r' R ,u, z' D) .
\label{eq:a28} 
\end{eqnarray} 

	These transformations are formal, but they
enable us to correctly drop the terms whose effect
is small.  Once these
terms are dropped, the equations reduce
to two dimensional equations,  which are more
easily handled numerically.
Although we have assumed that the dimensionless variables
are originally of order unity, it may be the case that they
differ substantially from unity.  However, an examination
of the various possible relevant  cases leads to the conviction
that all the important terms have been kept as
well as  other terms which are, perhaps, unimportant.
For example, the initial  value of $  B'_{\theta} $
is much smaller than unity. 
However, because of the shearing terms in equation 80
(the last term) $  B'_{\theta} $ grows to finite order
when $ t' $ becomes of order unity, so during the
later stages of the galactic disk, $  B'_{\theta} $ is of order unity.

	Many of the terms in 
equations 79 to 81 
 have an obvious significance.
The second term on the right hand side of the $  B'_{\theta} $
equation is the vertical decompression term present 
in the one dimensional simulation.  The 
first term is the radial decompression term, which only
becomes important when $ t' \approx 1 $, and the wrapping up
has made  the radial ambipolar diffusion important.

	Similarly, there is a $ z $ decompressional
term in the $ B'_r    $ equation, but, of course, no
radial decompression term.   There is a term representing the
effect of shear on the toroidal field in increasing the 
radial component, and a similar term resulting
from the action of shear on the $ B'_z $ component.
These shear terms would be small if $  B_{\theta} $ 
were of order of $ B_r $, or if $ B_z $ were much smaller
than $ B_r $, which is the case initially.  However, $ B_z $
 is increased by shear terms over the age of the disk,
to a value  considerably larger than its initial value.

	Equations 79 to 84 
 only contain derivatives with respect to
$ t', z' $ and $ u $,  and none with respect to
$ r' $.  Thus, $ r' $ is only a parameter in these
  differential equations.   Therefore, the components
of the magnetic field evolve independently of those 
at a different value of $ r' $ (to lowest order
in $ D/R $).  

	Further,  $ u $ does not occur explicitly
in the differential equations.  The initial conditions,
equations 85
do involve $ u $, and are periodic in it.  Therefore,
the magnetic field components remain periodic
in $ u $ for all $ t' $.

	Let us consider the behavior of such a solution,
periodic in $ u  $, in the neighborhood of the sun, $ r' = 1 $,
at fixed $ z' $, say $ z' = 0 $.   Transform the
solution back to $ r, \theta $ coordinates.  For fixed
$ r $ and $ t $, the solution is periodic in $ \theta, $
e.g. $ B_{\theta} (r, \theta ) = B_{\theta} (r, u+ \Omega(r) t)$.
Moreover, for fixed $ \theta $ we can write
\begin{eqnarray} 
u & = &  \theta - \Omega(r) t = \theta - \frac{\Delta r}{r}
\Omega_0 t \\ \nonumber 
&=& \theta - \frac{\Delta r }{r }t' \frac{R}{D}
 = \theta - \frac{\Delta r t'}{D}  , 
\label{eq:a29} 
\end{eqnarray}  
so 
\begin{equation}
\Delta r  = \theta - u \frac{D}{t'} .
\label{eq:a30} 
\end{equation}

	Thus, for fixed $ \theta, r  $ changes by an amount $ 
2 \pi D/ t' $ when $ u $ changes by its periodic length 
 $ 2 \pi $.  Since $ r' $ changes by a small amount 
$ \approx 2 \pi D/R t' $, we may ignore the dependence of 
the solution for the 
 components of $ {\bf  B} $ on $ r' $ and the components
of $ {\bf B } $ are nearly periodic in $ r $ ( at fixed $ \theta $ ).
However, because of the actual dependence
of the solution on $ r' $,  as a parameter in the equations, 
the amplitude and phase ( as well as the shape)
of the periodic solution do change slowly when
one goes a distance comparable with the radius of
the galaxy.

	Equations 79 to 84 were integrated numerically.
The details of the integration are discussed
in Howard(1995, 1996), where most of the results
are  presented. Initial conditions  were set by 
starting with a cosmic field before compression
into the disk, and then calculating the
resultant fields.  In this paper we present the results
for two initial cases.  Only the results for the integrations
at $ r =R $, the radius of the galactic solar orbit, are included.
The variation of  $ B_{\theta} $ as a function of $ u $, at 
$ z= 0 $, is plotted in figure 7 for the case that the
initial cosmic field was uniform.  The antisymmetry
in $ u $ is evident, and it is clear, after transforming the field
to be a  function of 
$ r $ as the independent variable by equation 87,
that   no Faraday rotation would be
produced by this field.  The same  result 
for the case when 
the initial cosmic  field was nonuniform is shown in figure 8.
[The initial cosmic field was chosen so after compression
into the disk the horizontal field was $ {\bf  B } =
 B_i [.5 + x ){\bf  \hat{x}} +  y {\bf  \hat{y}} ) $.]
The resulting saturated  field is not antisymmetric, and does not 
average out in $ u $.  It also would not average out
when transformed to be a function of $ r $, and {\em would}
produce a Faraday rotation.  The variation
of $ B_{\theta} $ with $ z $ at $ u = 0 $
at a  time of 9 gigayears, is shown in figure 9.
It has the parabolic shape found in section 3.
The variation of $ B_{\theta} $ with time  at the point 
$ u= 0, z = 0 $, is given in figure 10  for the two cases.  
The  results  are also  similar to those of figure 2
derived from the simple parabolic approximation.

\section{Ambipolar Diffusion in the Interstellar Medium}

	We now consider the averaged equations for
the magnetic field, taking into account the interstellar 
clouds.  The bulk of interstellar matter 
is in the form of diffuse clouds and
molecular clouds.  Because
the properties of the molecular clouds are
not very well known, we make the simplifying assumption
that essentially 
all the interstellar matter is in diffuse clouds,
with a small  amount of matter in the intercloud region.
In describing the clouds,  we make use of 
properties given by Spitzer(1968).
	
	We assume that all the clouds are
identical.  We further 
include the cosmic ray pressure, and the magnetic pressure,
in the intercloud
region, but neglect the pressure of the intercloud matter.
  Then
the cosmic rays and the magnetic fields are held in
the disk against their outward  pressures by the 
weight of the clouds in the gravitational field of
the stars. (See figure 3.)  

	Now the force due to the magnetic  and cosmic ray pressure
gradients  
is exerted only on the ionized matter in the clouds, 
while the gravitational force is exerted  mainly on the 
neutrals in the clouds,  since the fraction of ionization 
in the clouds is generally very low.  Thus, these contrary forces pull
the ions through the neutrals with ambipolar diffusion 
velocity $ v_D $. The frictional force between the
ions and neutrals is proportional to $ v_D $. 
By equating the magnetic plus the cosmic ray force to
the frictional force, we can  obtain 
the mean ambipolar velocity in the clouds.
Now,  we assume that the cosmic ray pressure 
$ p_R $ is related to the magnetic pressure $ B^2/8 \pi $
by the factor  $ \beta/\alpha $(Spitzer 1968).  We take $ \beta/\alpha $  
independent of time and space.  
This is plausible since when the magnetic field is
strong, we expect the  cosmic ray confinement
to be better and therefore the cosmic ray pressure
to be larger.

	 The mean vertical force per unit volume
produced by the magnetic field strength gradients
and cosmic ray pressure gradients is
\begin{equation}
F = - ( 1 + \beta/\alpha ) \frac{\partial B^2 / 8 \pi  }{ \partial z} .
\end{equation}
But this force is counterbalanced by the gravitational force on
neutrals  in the clouds,  which
occupy a fractional volume equal to the 
filling factor, $ f $, times the total volume.

	Thus,  the force per unit volume
on the ions in the clouds is
\begin{equation}
F_{cloud} = \frac{F}{f} = 
 - ( 1 + \beta/\alpha ) \frac{1}{f} 
\frac{\partial B^2/ 8 \pi  }{ \partial z} . 
\end{equation}
This force produces an ambipolar velocity, $ v_D $, 
  of the ions relative to the neutrals 
such that
\begin{equation}
F_{cloud} = n_i m^{*}  \nu v_D ,
\end{equation}
where $ m^{*} $ is the mean ion mass, $ \nu $ is the
effective ion--neutral collision rate for momentum 
transfer and $ n_i $ is the ion number density,  in
the clouds.  Now,  
\begin{equation}
\nu \approx n_c \frac{m_H}{m^{*}+m_H } <\sigma v > 
\end{equation}
where $ \sigma $ is the momentum transfer collision
cross section, and we assume that the neutrals are all
hydrogen with atomic mass $ m_H $.
Thus, we have 
\begin{equation}
F_{cloud} = n_i \frac{m_H m^{*} }{m^{*}+m_H } <\sigma v > n_c v_D 
\end{equation}

	If $ m^{*} >> m_H $, then the mass factor is $ m_H $.
Hence, if the ions are mostly singly charged carbon, 
the mass factor is $ \approx m_H $, while if they
are mostly hydrogen ions then it is $ m_H/2 $.
The ions are tied to the magnetic field 
lines, since the plasma is effectively infinitely 
conducting.  If there are several species of ions,  
 they all have the same cross field 
ambipolar velocity.  

	Thus, the ambipolar diffusion velocity in the  clouds
is
\begin{equation}
v_D = - \frac{1}{f} \frac{(1 + \beta/\alpha )\nabla (B^2/8 \pi)}
{n_i m^{*}_{eff} < \sigma v > n_c } ,
\label{eq:93} 
\end{equation}
where $ m^{*}_{eff} = <m^{*} m_H/(m^{*} + m_H)> $ 
is the effective mass averaged over ion species,
and $ \sigma $ is the ion--neutral cross section.

	The velocity $ v_D $ is the mean velocity 
of an ion inside a cloud.  It is also the velocity of
a given line of force.  Now, the ions are continually
recombining and being replaced by other ions, 
so it is  actually the motion of the
magnetic  lines of force that has significance.
Further, the clouds themselves have a short life 
compared to the age of the disk.  They
collide with other clouds 
every $ 10^{ 7} $ years or so, and then quickly reform.
After the collision the cloud material is dispersed, 
but because of its high conductivity,  it stays connected
to the  same lines of force.  After the cloud
reforms the lines of force are still  connected to the same
mass.  The lines then continue to move in the reformed 
cloud,  again in the opposite direction to  the field gradient.
As a consequence, during each cloud lifetime,
 the lines of force pass through  a certain  amount of mass
before the cloud is destroyed. 

	 The amount of mass 
passed during a cloud life time, by a single flux tube of diameter $ d $
is $ 2 v_D t_c  a d \rho_c $ where $ t_c $ is the cloud life time,
$ 2 a $   is the cloud diameter, 
 and $ \rho_c $ is the cloud density.   However,
because of our assumption that the bulk of the matter is
in these diffuse clouds, it must be the case that
there is only a short time in between the destruction 
of one cloud, and its reformation.  Thus,
during a time $ t $, the field passes through 
$ \approx t/t_c $ successive clouds.  If we consider
a  length of a line of force $ L $,  at any one time
it passes through $ f L /2 a $ different clouds. 
Then, in the time $ t $, the amount of matter
passed by this length of a given tube of force
(of diameter $  d $ ) is
\begin{equation}
\Delta M = 2 a d  v_D t_c \rho_c \frac{L}{2a} \frac{ t}{t_c}
=L f \rho_c d v_D t.
\end{equation}
But $ f \rho_c   $, is the mean density, $ \rho $,
  of interstellar matter.
Thus, if we define $ \Delta x $ as  the average effective distance
 that this magnetic tube
moves through the disk by
\begin{equation}
\Delta M = \rho \Delta x L d ,
\end{equation}
we get 
\begin{equation}
d \rho \Delta x L = L d \rho v_D t ,
\end{equation}
or 
\begin{equation}
\frac{\Delta x}{t} = v_D .
\end{equation}

	This 
effective {\em velocity},  averaged over many cloud lifetimes,
is the  only  velocity for the magnetic field lines  that makes sense.
It is equal to  the  velocity, $ v_D $, of the ions
in the cloud material.

	So far, we have made the simplifying assumption
that  the clouds
are stationary,  which is of course not the case. 
As the clouds move through the interstellar medium  they
stretch the lines, and additional tension forces
and ambipolar velocities arise.  However,
these velocities are always directed toward the
mean position of the cloud material,  and thus they
tend to average to zero.  The ambipolar velocity which
 we have calculated above, under the assumption of
stationary clouds, actually gives the rate of displacement of the
lines of force relative to the mean position
of the cloud.  It is a secular velocity, and it is the only
velocity which really  counts.

	We assume that in the diffuse interstellar clouds
only carbon is ionized, so that $ m*_{eff} = m_{H} $.
We choose $ < \sigma v > = 
2 \times 10^{ -9} \mbox{cm}^2/ \mbox{sec} $(Spitzer 1978).
 If the filling factor of the clouds $ f = 0.1 $, and
if the mean interstellar density of hydrogen is
 $ n_0 \approx 1.0  / \mbox{cm}^3 $,   then
the density in the clouds is $ n_c \approx 10 / \mbox{cm}^3 $.
The cosmic abundance of carbon is $ 3 \times 10^{ -4} $(Allen 1963).
 We assume that this abundance is depleted by $ \zeta_0 \approx 0.2 $,
(Spitzer 1978), the rest of the carbon being locked up in grains. 
 We further take
$ \beta/\alpha = 2 $.  Then  comparing equation~\ref{eq:40} with
equation~\ref{eq:93} gives 
\begin{equation}
K = \frac{( 1 + \beta / \alpha) }{n_i m*_{eff} < \sigma v > n_0} 
= 1.5 \times 10^{ 36} , 
\end{equation}
in cgs units.  Taking $ D = 100  $ pc, we get from equation~\ref{eq:54}
\begin{equation}
B_0 = 2.85 \times 10^{ -6} \mbox{gauss} .
\end{equation} 

We take $ t_0 = 3 \times 10^{ 9} $ years.   We see that if 
$ t' = 3 $,  corresponding to an age for the galactic disk
of $ 9 \times 10^{ 9} $ years,  then 
the present value of the magnetic field
from figure 8,  should be $ 1.5  \times 10^{ -6} $
gauss.

This value depends on our assumptions concerning the properties
of the clouds.  In particular, if the hydrogen in the
interstellar clouds is partly ionized by low energy
cosmic rays penetrating the clouds, then the ambipolar diffusion
will be slower, and $ K $ will be smaller, leading to 
a larger value for $ B_0 $.  The consequence of this
is, that the initially dimensionless magnetic field 
$ B'_r $ will be smaller, so that it would take longer
to reach saturation.  However,  the saturated value
will be larger.

\section{Conclusion}

	We have assumed that there
was a cosmic magnetic field present before
the galaxy  formed.  On the basis of this
hypothesis we have constructed a simplified model
of the galactic disk in order to investigate
how the magnetic field 
would evolve, and what it would look
like at present.  The two essential ingredients
of this model are:  the differential rotation
of the interstellar medium,  and the motion
of the field lines through the interstellar medium produced
by the ambipolar diffusion of the  ionized component 
of the plasma driven through the neutrals by magnetic pressure and 
cosmic ray pressure gradients. The effect
of turbulent motions is assumed to average out.
 No large scale
mean field dynamo action on the magnetic
 field was included in this model. (This model
differs from that of Piddington in that the magnetic
 field is too weak to effect the galactic rotation 
of the interstellar medium,  and  also, ambipolar diffusion is 
included.)

	The consequences of the model 
were investigated  by an approximate  analytical
calculation in Section 2,  and by  more detailed
 numerical 
simulations in Section 3  and 4.  These simulations
 confirmed the results of the approximate  analysis
of section 2.

	The basic results are:

	(1) To first approximation,
 the magnetic field evolves
locally following a rotating fluid
element.  It first grows by stretching
the radial component of the magnetic field into
the toroidal direction.  When the field becomes
strong enough, the line commences to shorten 
because of the vertical motions  produced by
ambipolar diffusion.  This reduces the radial
component, and therefore the stretching. 
After a certain time,  the field strength
saturates and starts to decrease as the
 reciprocal square root of time.
This asymptotic behavior is determined
only by the ambipolar diffusion properties
of the clouds.  Thus, the field strength everywhere 
approaches the same value at a given time. 
At the present time,  this value is estimated to
be in the range of a few microgauss.  This saturated  value
is independent of the initial value which  the
field had  when the disk  first formed,  provided that the
initial value of the field strength 
is greater than $ 10^{ -8} $  gauss.
The extent of each magnetic field line 
in the toroidal direction also saturates,
but its length in the disk {\em does}  depend on the 
initial value.

	(2) The direction of the toroidal field 
in any given fluid element depends on the sign
of the initial radial component.  
Since this sign varies with position, and since
differential rotation mixes these positions,
it turns out that the resulting toroidal field 
varies rapidly with radius along a fixed radial direction.
The toroidal field changes direction on a scale of a hundred
parsecs.  Because the saturated field strength 
is nearly constant in magnitude, the toroidal field
strength as a function of $   r $ at fixed $ \theta $ 
varies as 
a square wave.  However, the lobes of this square 
wave need not be equal since the regions of one initial
sign of $ B_r $ may be larger than those of the other
sign.   In this case, the model predicts a toroidal  field that would
produce  a net Faraday rotation in radio sources such
as pulsars or polarized 
 extragalactic radio sources,  in spite of its rapid variation in sign.

	It is the prevailing belief that the 
galactic magnetic field does not reverse in radius on small
scales.  In fact, this belief is grounded in an analysis of 
Faraday rotation measures of pulsars(Hamilton and Lynn 1987). 
 In analyzing these
rotation measure Rand and Kulkarni(1988) employed 
 various simple  models of the 
galactic field.  In every one of these models,  the
magnetic field reversed only on large scales of
order one kiloparsec.  These models, which only
allowed a slow variation of $ B $, led to results that were
consistent with the observations, and thus supported the
general belief in reversal only on large scales.
However,  when this analysis was carried out,
 there was no apparent reason 
reason not to consider a model in which the field varied rapidly in
radius on scales of order of a hundred parsecs,
although in hindsight it could have been done.
 However,
 such a model
was not included in the analysis of the rotation measures,
and thus it was not tested.  Therefore, at present,  there is no reason
to exclude such a model.  

	In short, the prevailing
belief that the galactic  magnetic field is of  constant sign
on a large scale,  actually resulted from the
assumption that the field was a large scale field, and
from the consistency of this assumed model with observations.  
Therefore, this
 conclusion 
has not been rigorously demonstrated. A magnetic field, 
such as that  arrived at from  our model, which has the rapid
variation of the field with radius, would lead to
fluctuations from the mean.
Indeed,  such fluctuations are in the data and
are attributed to a general isotropic random
magnetic field in addition to the mean field.

	We feel the the non uniform square wave could probably
fit the observations equally as well as the other models,
so that our model can also be shown to be consistent with
 observations.
We have not yet  demonstrated quantitative
consistency with observations,   but we hope
to carry out this task in the future.

	(3)  The  magnetic
field observed in our galaxy and in other
galaxies should actually be the average of the true detailed magnetic field
averaged over regions in space larger than those
regions over which we find our field to vary
(several hundred parsecs at least).  Thus,
under averaging the field of  our model would actually
appear as an 
axisymmetric toroidal magnetic field.  An analysis of the various
galactic magnetic fields has made the {\em ansatz} 
that the origin  of the field can be distinguished as
primordial if it has  
bisymmetric symmetry, and as due to a dynamo if
it has
axisymmetry(Sofue et al. 1990). 
(It must be borne in mind that the actual  magnetic field is perturbed by
the spiral arms, so that it  is parallel
to the arms in the region of the spiral arms and 
toroidal in the region  in between the arms as described 
in the introduction.)    On the basis of our
analysis, we conclude that this method of 
distinguishing the origin  is not  valid,
and could  lead to incorrect conclusions concerning the
origin of the galactic magnetic fields.

	4.  Parker(1968, 1973a,b)   and others(Ruzmaikin et al 1988) 
 have 
put forth three objections to a 
 primordial origin.  These objections  are:
 (a) A primordial 
field would be tightly wrapped up,  contrary to observations.
(b) Such a field would be expelled by ambipolar diffusion
or turbulent diffusion.
 (c) There is no known mechanism to produce a
large scale primordial field in the early universe.

	  On the basis of  our model we can counter the first
two  objections (a) and (b). Our model does
indeed lead to a tightly wrapped spiral magnetic field.
However, 
when averaged over a sufficiently large scale,  as is
done automatically by observations, 
 the resulting field should actually
appear to be axisymmetric and  azimuthal.  
This averaged field is, thus,  not in disagreement with
observations.  In addition, if the
large scale field is not uniform, then the 
field after saturation
by ambipolar diffusion creates larger regions
of one sign then of the other sign 
and the  averaged field is not zero.   Thus,  objection
(a) does not defeat our model for a
field of primordial origin.

	With respect to objection (b) our model predicts field
lines which thread through the disk, entering
on one edge and leaving on the other.  Thus, it is 
impossible for a vertical ambipolar motion or 
turbulent mixing to
expel lines of force.  This result counters  objection (b).  

	The third objection is still open to debate
and we do not discuss it.
	
	The two observations that should test our model are:
The magnetic field in other galaxies should be observed
to be toroidal  in between the arms. A reanalysis
of the pulsar rotation measures should fit our model
without very large extra fluctuations.  These fluctuations
should be accounted for by the rapid reversals of the
toroidal field predicted by our model.

	To summarize, we have shown that a
careful analysis of the evolution of a primordial
field throws new light on the way one should 
view the primordial field hypothesis.

\acknowledgments
The authors are grateful for a helpful discussions
with Steve Cowley and Ellen Zweibel. 
The work was  supported by the  National Science
Foundation  Grant AST 91-21847
and by  NASA's astrophysical program
under  grant  NAG5 2796. ACH was supported by  an
NSF minority fellowship and an AT\&T Bell labs 
Cooperative Research Fellowship.

\clearpage

\appendix
\section{Appendix A}

	In this appendix we give the derivation of 
 the solution given in equations 28-30,
to the differential equations  (24), (25),  and (27) of 
section 2.  Let us drop the $ \theta $ subscript
in $ B_{\theta} $.  
Then the equations to be solved are
\begin{eqnarray}
\frac{d B_r}{dt} & = & - \frac{v_D}{D} B_r   \\ \nonumber 
\frac{d  B}{d t} & = & -  \frac{v_D}{D} B  - \Omega B_r  \\ \nonumber  
\frac{v_D }{D} v_D & = &  k  B^2  .
\end{eqnarray}

	Define $ \xi $ by
\begin{equation}
\frac{d \xi }{d t} = \frac{v_D}{D} = k B^2 .
\end{equation} 
Then
\begin{equation} 
\frac{d B_r }{d \xi } =   - B_r ,
\end{equation}
or 
\begin{equation}
B_r = B_1 e^{ - \xi } ,
\label{eq:4app}
\end{equation}
and 
\begin{equation}
\frac{d B}{d \xi } = - B - \frac{\Omega }{k  B^2 } =
 - B - \frac{C}{B^2}e^{ -\xi } ,
\end{equation}
where $ C= \Omega B_1 /k  $ is a constant.

Multiplying by $ e^{ \xi } $, we get
\begin{equation}
\frac{d}{d \xi } (B e^{ \xi } ) = \frac{C}{B^2} = 
\frac{C e^{2 \xi}}{(B e^{ \xi} )^2} .
\end{equation}

Now, $ B $ is negative.  Let $ y = - B e^{ \xi} $. Then
\begin{equation}
\frac{d y}{d \xi } = \frac{C e^{2 \xi} }{y^2} .
\end{equation}
Integrating and assuming the initial value of $ y $ is small
we get
\begin{equation}
\frac{y^3}{3} = \frac{C}{2} ( e^{2 \xi} - 1)  ,
\end{equation}
or
\begin{equation}
y  = (\frac{3 C}{2})^{1/3} (e^{2 \xi} -1 )^{1/3}  .
\label{eq:9ap}
\end{equation}
 
	Now, 
\begin{equation}
e^{ 2 \xi} \frac{d \xi }{d \tau } = k  B^2 e^{2 \xi} .
= k y^2 
\end{equation}

	Next, let $ e^{ \xi} = \eta $.
Then
\begin{equation}
\frac{d \eta }{d t  } =
 \frac{2 k}{C} (\frac{3C}{2})^{2/3} ( \eta - 1 )^{2/3} ,
\end{equation}
or
\begin{equation}
\frac{d \eta }{(\eta - 1 )^{2/3}} = 2 k ( \frac{3}{2}C)^{2/3} d t  .
\end{equation}
Integrating this equation we have  
\begin{equation}
(\eta - 1 )^{1/3} = \frac{2}{3  } ( \frac{3}{2} C)^{2/3} t
\end{equation}
Now expanding this equation and making use of the definition of
$ \eta $ we have
\begin{equation}
e^{2 \xi} = 1 + \frac{2}{3} C^2 k^3 t^3 t^3 .
\label{eq:14app}
\end{equation}

The toroidal field $ B = y e^{ - \xi} $, and $ y $ is given
by equation~\ref{eq:9ap} so 
\begin{equation}
B^2 = y^2 e^{ -2 \xi} = (\frac{3 C}{2})^{2/3} 
\frac{(e^{2 \xi} - 1 )^{2/3}}{e^{2 \xi } }  ,
\end{equation}
or 
\begin{equation}
B^2= \frac{C^2 k^2 t^2}
{1+ (2/3) C^2 k^3 t^3}  .
\end{equation}
Restoring the definition  $ C $, and setting $ v_{D1} = k B_1^2 $ 
 we get equation
(29) of section II for $ B_{\theta} $.
 Equation (28)  for $ B_r $ is obtained from equation~\ref{eq:4app}
 and equation~\ref{eq:14app}.

\clearpage

\section{Appendix B}

In this appendix we give the full equations for
$ d B'_r /dt , d B'_{\theta} / dt $  and
$ d B'_z /dt $ including all the terms in $ R/D $.
They result from the transformation of equations  64 to 69
to dimensionless variables by making use of equations 70 to 77.

  For convenience, the terms are listed in exactly
the same order as in the original dimensional equations.
The resulting equations are
\begin{eqnarray} 
\frac{\partial B'_r }{\partial t'} & = &
   \frac{B'_{\theta} }{r'} \frac{\partial w'_r }{\partial z'}  
        -B'_z \frac{\partial w'_r }{\partial z'}    \\ \nonumber  
 &  &  
 - \left(\frac{D}{R}\right)   w'_r  \frac{\partial B'_r }{\partial r'}  
        -\left(\frac{D}{R}\right)^2 \frac{w'_{\theta} }{r'} 
          \frac{\partial B'_r }{\partial u} 
             -  t' \frac{w'_r }{r'^2}\frac{\partial B'_r }{\partial u}   
- w'_z \frac{\partial B'_r }{\partial z'}  \\ \nonumber  
& &
   - \left(\frac{D}{R}\right) \frac{w'_r }{r'} B'_r 
         - \left(\frac{D}{R}\right)^2 \frac{B'_r }{r'} 
              \frac{\partial w'_{\theta} }{\partial u}  
                   - B'_r \frac{\partial w'_z }{\partial z'}  
\end{eqnarray}
\vskip .2 in
\begin{eqnarray}
\frac{\partial B_{\theta} }{\partial t'} & = & =
 \left(\frac{D}{R}\right) ^3 B'_r \frac{\partial w_{\theta} }{\partial r'} 
 + \left(\frac{D}{R}\right) ^2 \frac{B'_r }{r'^2} 
     \frac{\partial w'_{\theta} }{\partial u} t'   \\ \nonumber  
&  & 
 + \left(\frac{D}{R}\right)  B'_z \frac{\partial w'_{\theta} }{\partial z'} 
 - \left(\frac{D}{R}\right)  w'_r \frac{\partial B_{\theta} }{\partial r'} 
 - \left(\frac{D}{R}\right) ^2 \frac{w'_{\theta} }{r'}
       \frac{\partial B'_{\theta} }{\partial u} 
  - \frac{t'}{r'^2} w'_r \frac{\partial B'_{\theta} }{\partial u} 
\frac{\partial B_{\theta} }{\partial r'}    \\ \nonumber  
&  & 
   - w'_z \frac{\partial B'_{\theta} }{\partial z'} 
  -\frac{B'_r }{r'}  \\ \nonumber 
&  &
   - \left(\frac{D}{R}\right)  B'_{\theta} 
      \frac{\partial w'_r }{\partial r'} 
   - \frac{t'}{r'^2} B'_{\theta} \frac{\partial w'_r }{\partial r'} 
   - B'_{\theta} \frac{\partial w'_z }{\partial z'}  
\end{eqnarray}
\vskip .2 in
\begin{eqnarray}
\frac{\partial B'_z }{\partial t'} & = & 
   \left(\frac{D}{R}\right)  B'_r \frac{\partial w'_z }{\partial r'}   
   + \frac{B'_{\theta} }{r'} \frac{\partial w'_z }{\partial u}  
  + \frac{t'}{r'^2} B'_r \frac{\partial w'_z }{\partial u}   \\ \nonumber  
&  & 
   - \left(\frac{D}{R}\right)  w'_r \frac{\partial B'_z }{\partial r'}  
   - \left(\frac{D}{R}\right) ^2 \frac{w'_{\theta} }{r'}
        \frac{\partial B'_z }{\partial u}
   - \frac{t'}{r'^2} w'_r \frac{\partial B'_z }{\partial u}  \\ \nonumber 
&  & 
    - w'_z \frac{\partial B'_z }{\partial z'}  
    - \left(\frac{D}{R}\right)  B'_z 
        \frac{\partial w'_r }{\partial r'}  \\ \nonumber 
&  & 
     - \frac{t'}{r'^2} B'_z \frac{\partial w'_r }{\partial u} 
     - \left(\frac{D}{R}\right)  \frac{B'_z }{r'} w'_r 
    - \left(\frac{D}{R}\right) ^2 \frac{B'_z }{r'}
       \frac{\partial w'_{\theta} }{\partial u}  
\end{eqnarray}
\vskip .2 in
 
	The equations for the components of $ {\bf  w } $ 
are:
\begin{eqnarray}
w'_r & = & 
- \frac{t'}{r'^2} \frac{\partial B'^2 }{\partial u}  
            +\frac{\partial B' }{\partial r'}  \\ \nonumber  
w'_{\theta} & = & 
  - \frac{1}{r'} \frac{\partial B'^2 }{\partial u}  \\ \nonumber  
w'_z  & = &
- \frac{\partial B'^2 }{\partial z'}  
\end{eqnarray} 

In these equations $ B'^2 $ stands for
\begin{equation}
B^{'2} = B^{'2}_{\theta} + \left(\frac{D}{R}\right)^2 (B^{'2}_r + B^{'2}_z )
\end{equation}
We do not expand the expression for the components of $ {\bf  w'} $
out fully.

\clearpage

\begin{center}
{\bf FIGURE CAPTIONS}
\end{center}

\begin{description}

\item[Fig. 1] (a) The protogalaxy before initial 
collapse, showing a uniform cosmic field threading it.
The magnetic field makes an angle $ \alpha $ with the
rotation axis $ \Omega $.  
(b) The magnetic field after the protogalaxy has collapsed
to the present size of the galaxy but before
the galactic disk has formed.  The lines remain
threaded through the galactic plasma and connected
to the rest of the cosmos.  
(c) The magnetic field lines after collapse to
the galactic disk.  Some of the lines such as
a and d will escape by ambipolar diffusion and
turbulent mixing.  The lines $  b $  and $ c $ cannot
escape.
(d) The vertical ambipolar diffusion velocities
act in opposite directions on a line of force
to shorten its horizontal extent in the disk.
They cannot remove it from the disk.

\item[Fig. 2] The variation of the magnetic field strength
of a line of force for different initial conditions.
The field strength at first grows by stretching, 
and later decreases because of ambipolar diffusion.
The saturated state of the lines is the same,
independent of its initial field strength.
The initial values of B  in microgauss 
were, .01,.03, .1, .3, 1.0, and 3. 

\item[Fig. 3] The interstellar medium consists
of dense clouds shown as spheres.  A magnetic field
line is shown which threads several clouds.
The bowing up of the line in between the clouds
is caused by cosmic ray and magnetic pressure in
the intercloud medium. These pressures lead to 
a upward force on the neutrals in the cloud.  The cloud is held down
by gravitational force on the neutrals.
The action of these two forces  produces
a motion of the ions through the cloud,
ambipolar diffusion. The right hand
end of the line is connected to the external
cosmos. When the end of the line
escapes from the last cloud by ambipolar diffusion, 
this part of the line will be expelled into the
external world, and the part of the line 
in the disk will become shorter.

\item[Fig. 4]The lines of force for a nonuniform
magnetic field after compression into the disk.
To preserve zero divergence the lines of force
must fan out in the direction of weaker field.
It is seen that the region of weaker
field, where $ B_r $ is negative, is actually
more extensive than the region where $ B_r $  is positive.
Thus, after the field everywhere reaches a saturated
field strength this weaker region will lead
to a region of larger area, and the final mean magnetic field will
be dominated by it. (Because the saturated
value is independent of the initial field strength,
the weakness of the initial field will 
not be important,  and only its more extensive area will be important.)

\item[Fig.5] The variation of the magnetic field 
strength with $ z $ for different times
as a result of the one dimensional
calculation.  At first the strength grows
by stretching, and afterwards decreases
by ambipolar diffusion.  The value is negative
because the initial radial field was positive,
and the differential rotation sweeps it backwards.
The magnetic field is plotted in dimensionless units.
The dashed curves correspond to the times
during which the field strength is increasing
while the solid lines correspond to the decreasing
phase.  The convergence of the curves for the larger
times is evident. 

\item[Fig. 6]  The variation of the magnetic field with
time for initial conditions $ B_r = B_i \cos \theta $ 
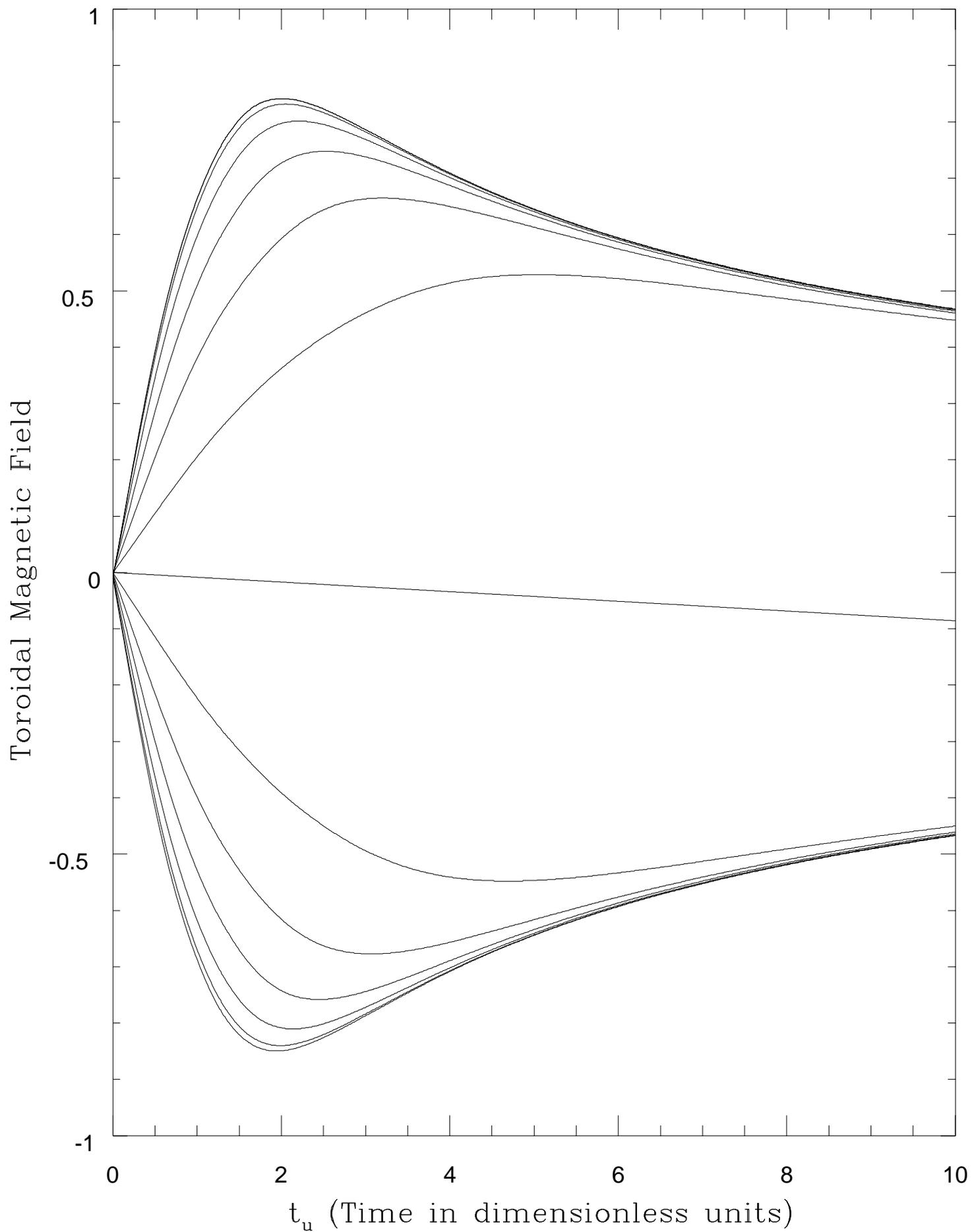
for $ \theta $ equal 0 to 180 degrees in 15 degree increments,
where $ B_i = .01 $  in dimensionless units.  These curves
are derived from the one dimensional calculation.  The 
convergence of the curves at large $ t $ parallels the same
convergence in figure 2, and the curves for equal
initial conditions  agree quantitatively. (The values of
the initial conditions in microgauss are $ 0.0285
 \times \cos \theta + 0.000285 =  0.0288, 0.0250, 0.0205,
0.0146, 0.0077 $, etc.  )

\item[Fig. 7]  The results of the two dimensional
numerical simulation,  for the case of an initial
uniform magnetic field,  $ B_{\theta} $ is
shown as function of $ u $ at $ z = 0 $ for different
times.  That the integral of $ B_{\theta}$ 
is zero is evident.  This field 
will not lead to any  Faraday rotation.
The labeling of the curves is the same as in figure 5.

\item[Fig. 8] The results of the two dimensional
numerical simulation,  for the case of an initial
non uniform magnetic field. $ B_{\theta} $ is
shown as function of $ u $ at $ z = 0 $ for different
times.  The regions of positive $ B_{\theta} $
are broader than those of negative $ B _{\theta} $, so
that the integral does not vanish, and this field 
{\em will }  lead to a non zero Faraday rotation
The labeling of the curves is the same as in figure 5.

\item[Fig. 9] The results of the two dimensional
numerical simulation,  for the case of an initial
non uniform magnetic field.  $ B_{\theta} $ is
shown as function of $ z $ at $ u= 0 $ for different
times.  The labeling of the curves is the same as in figure 5.
The similarity to figure 5 is noteworthy.

\item[Fig. 10] The results of the two dimensional
numerical simulation,  for the both  cases.  $ B_{\theta} $ is
shown as function of $ t $ at $ u= 0, z=0 $.
The behavior is quite similar to figure 2
for the one dimensional simulation.

\end{description}

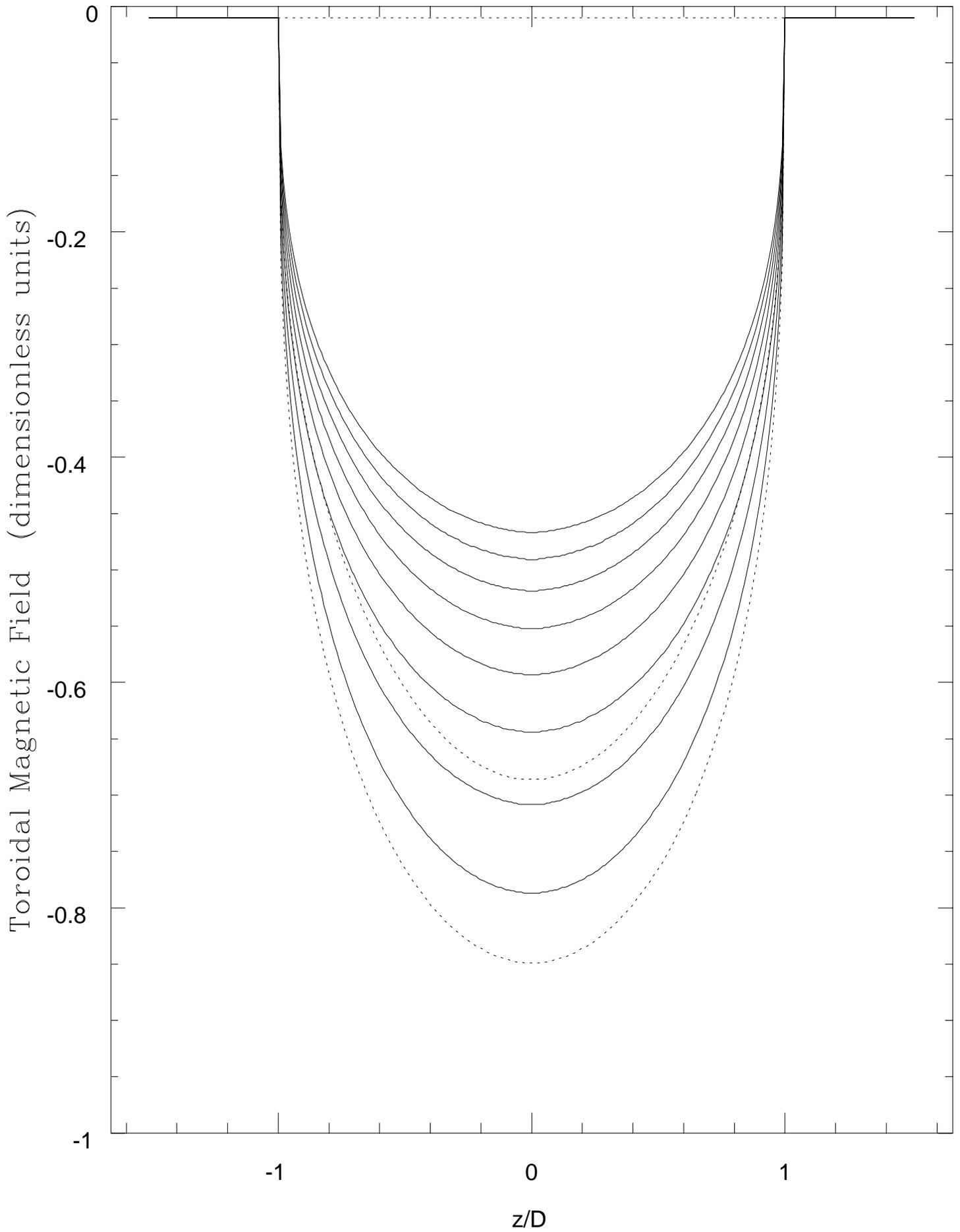
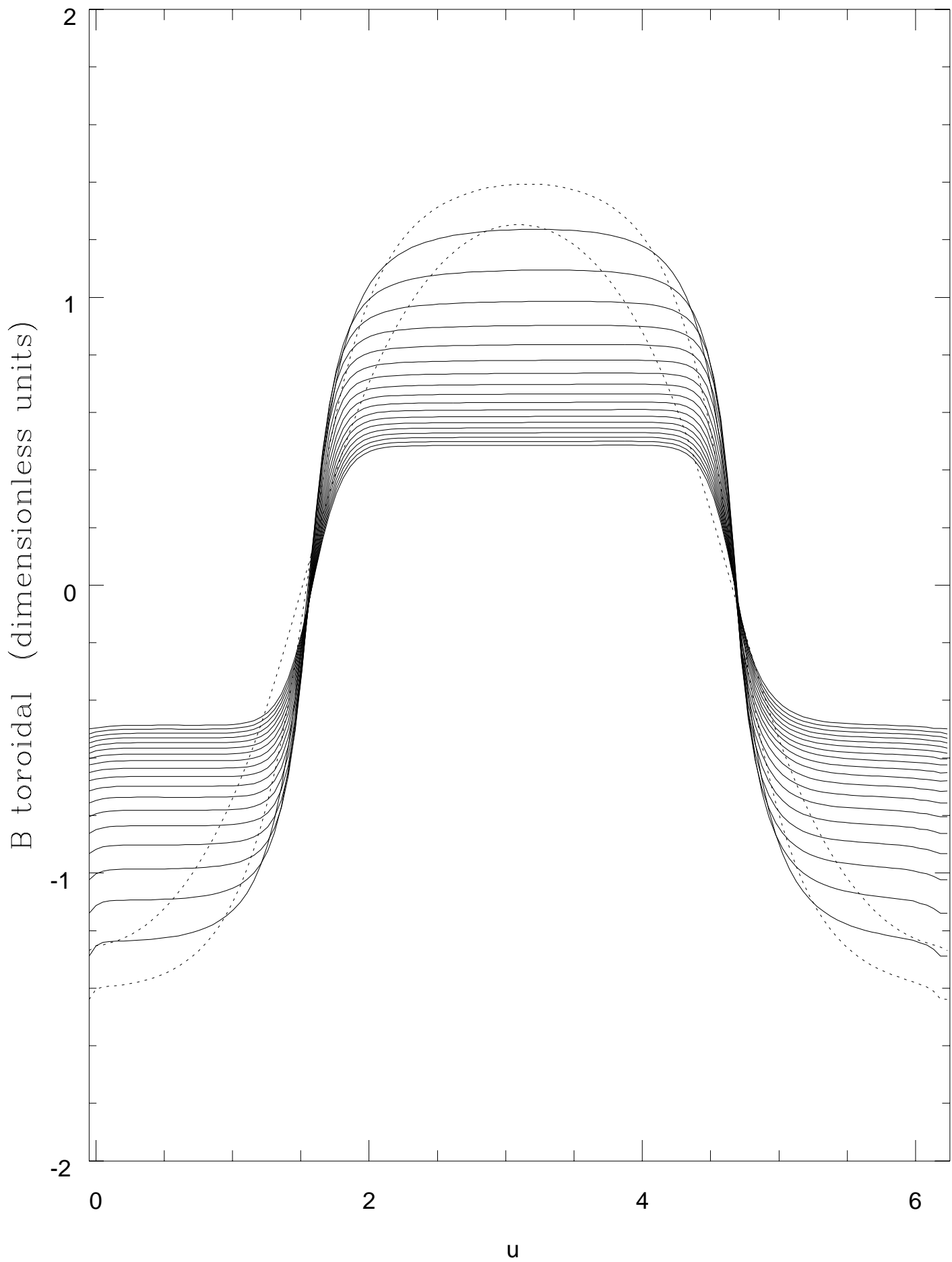
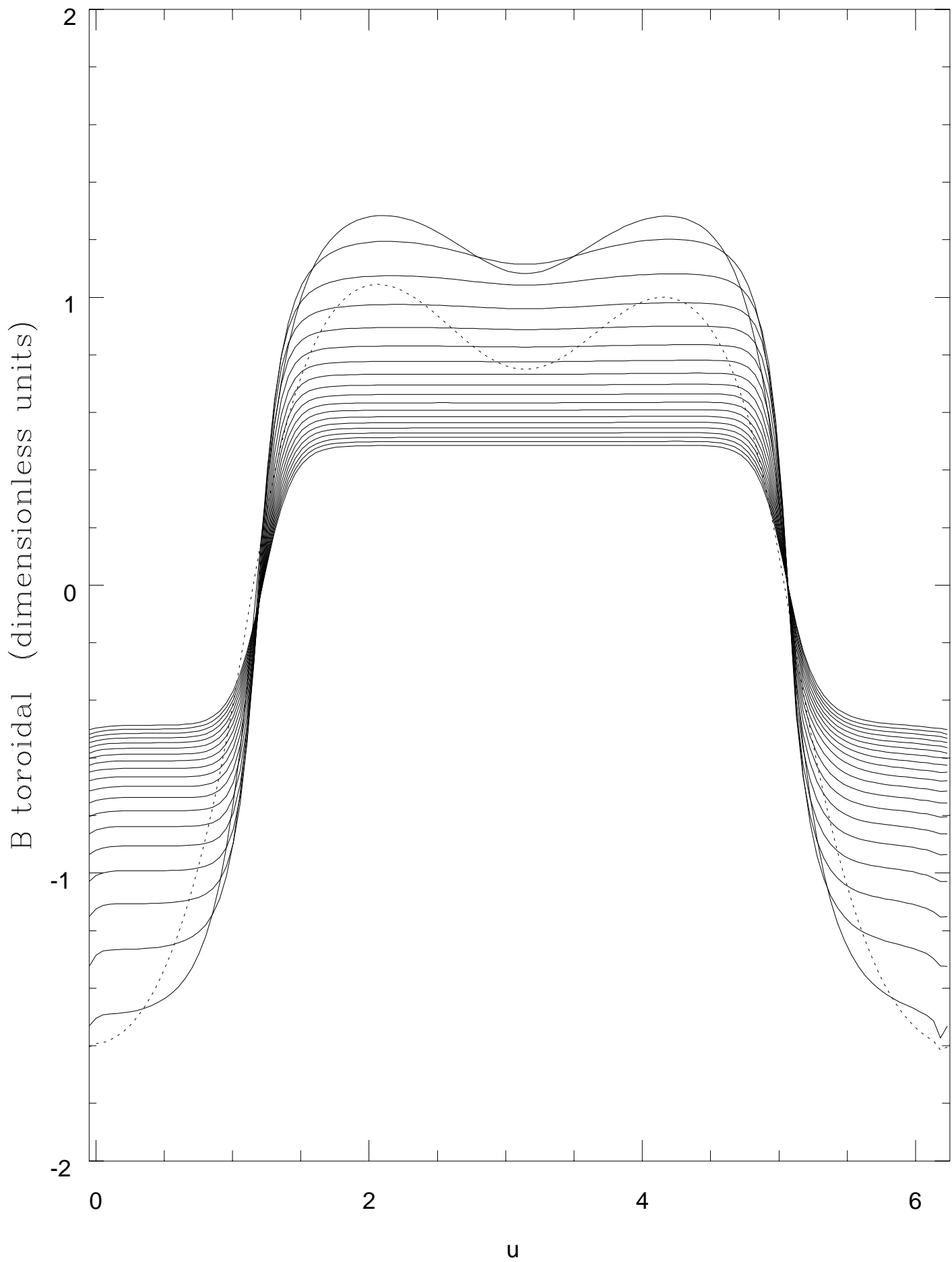
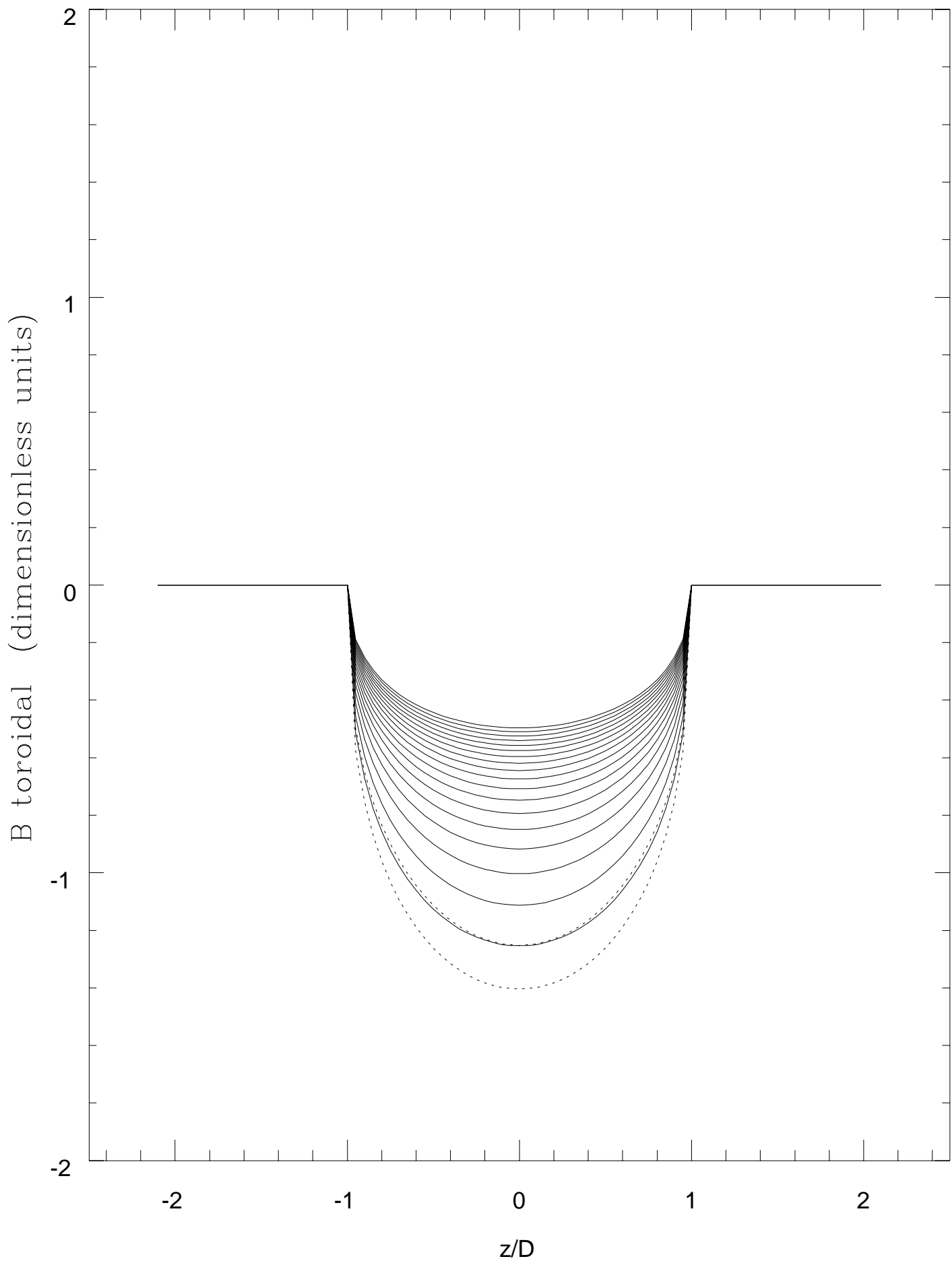
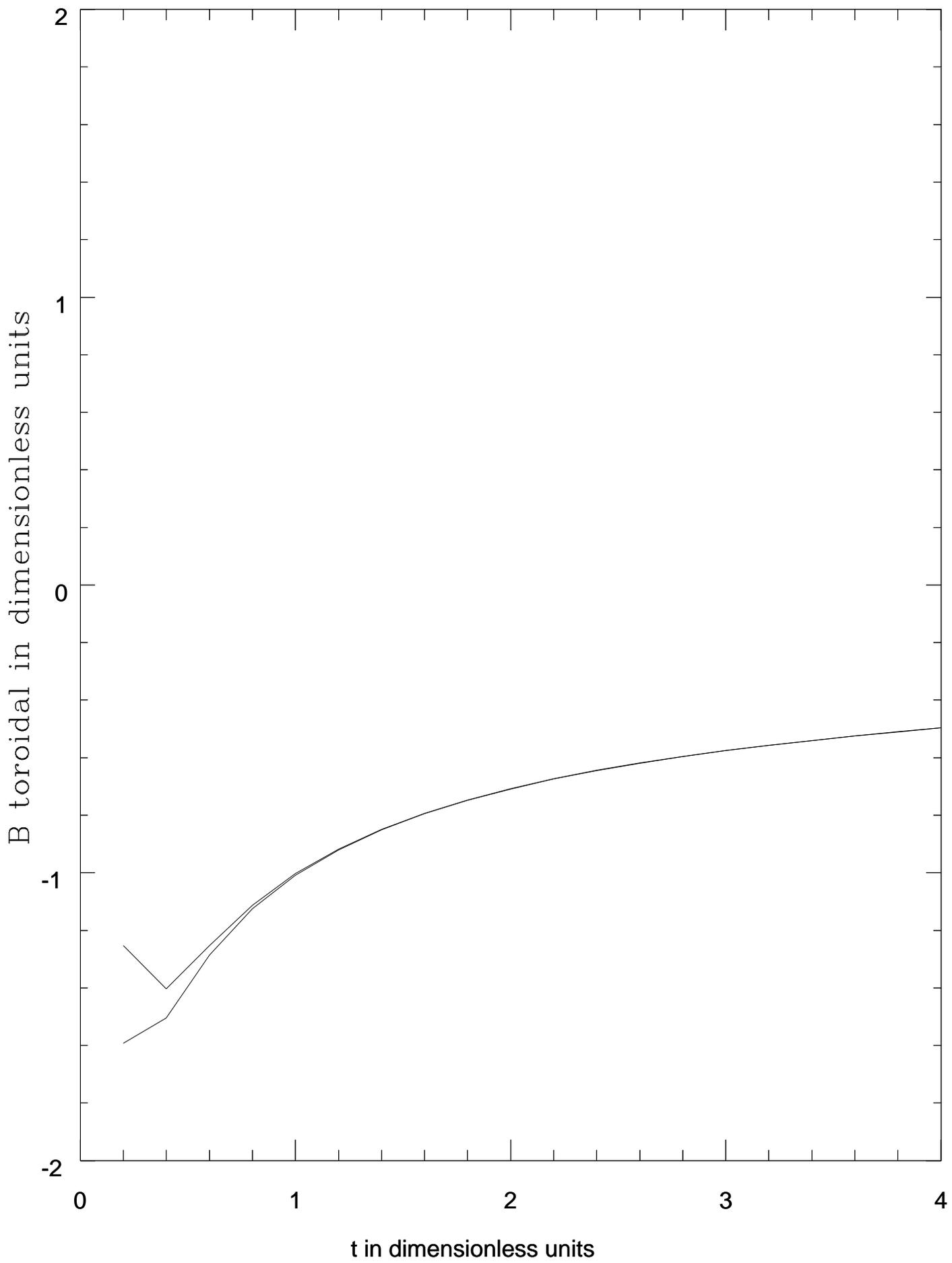
\end{document}